\documentstyle[preprint,aps,rsfs]{revtex}

\title{Transverse wavevector dependent and frequency dependent
       dielectric function, magnetic permittivity and generalized
       conductivity of interaction site fluids. MD calculations \\
       for the TIP4P water
      \vspace{6pt}}

\author{{\sc Igor~P.~Omelyan \vspace{9pt}}}

\address{Institute for Condensed Matter Physics,
         the National Ukrainian Academy of Sciences,
         1~Svientsitsky St., UA-290011 Lviv, Ukraine \thanks
        {E-mail: nep@icmp.lviv.ua}}

\newcommand{\bms}[1]{\mbox{\boldmath $#1$}}
\newcommand{\bvs}[1]{\mbox{\scriptsize\boldmath $#1$}}
\newcommand{\scs}[1]{_{\stackrel{\ }{#1}}}

\begin{document}

\maketitle

\vspace{12pt}

\begin{abstract}
It is shown that the dielectric properties of interaction site models
of polar fluids can be investigated in computer experiment using not
only the charge fluctuations but also correlations corresponding to
a current of moving charges. This current can be associated with a
generalized dynamical polarization or separated into electric and
magnetic components. The first approach deals with the dielectric
permittivity related to a generalized conductivity, whereas the
second one leads to the functions describing polarization and
magnetization fluctuations separately. The last way is only the
source to calculate the magnetic susceptibility for a system of
interaction sites. The transverse wavevector- and frequency-dependent
dielectric functions and magnetic susceptibility are evaluated for the
TIP4P water model in a very wide scale of wavelengths and frequencies
using molecular dynamics simulations. We demonstrate that the transverse
part of the dielectric functions may differ drastically from their
longitudinal component. A relationship between the two approaches
is discussed and the limiting transition to the static dielectric
constant in the infinite-wavelength regime is analyzed. The
propagation of transverse electromagnetic waves in the TIP4P
water is also considered.
\end{abstract}

\newpage
\section{Motivation}

In a recent paper [1], a computer adapted fluctuation formula suitable
for self-consistent calculations of the longitudinal wavevector- and
frequency-dependent dielectric function $\varepsilon \scs{\rm L}(k,
\omega)$ for interaction site models (ISMs) of polar systems has been
proposed. As a result, a detailed analysis of this function in the entire
wavelength and time scales was carried out for the TIP4P water model using
molecular dynamics (MD). It was shown that the choice of microscopic
variables for the operator $\bms{\hat P}$ of polarization density plays
an important role in a correct reproduction of dielectric properties,
and only true microscopic variables, which explicitly take into account
the charge distribution within molecules, can be used to determine the
frequency dependence of the dielectric permittivity of ISMs at arbitrary
wavenumbers.

It is a common practice to investigate the dielectric properties of ISMs on
the basis of charge fluctuations [1--3]. In the infinite-wavelength limit
$k \to 0$, these fluctuations reduce to the well-known longitudinal dipole
moment correlations, which are usually considered in computer experiment to
obtain $\varepsilon \scs{\rm L}(k,\omega)$ at zero and small wavevector
values [4--12]. The transverse dipole moment fluctuations are used sometimes
[5, 9, 10--12] for treating the transverse component $\varepsilon \scs{\rm
T}(k,\omega)$ of the dielectric permittivity. However, such an approach,
being exact for point dipole systems [13, 14], cannot be applied to
calculations of the genuine transverse dielectric function of ISMs
at nonzero wavenumbers.

The problem of computations of $\varepsilon \scs{\rm T}(k,\omega)$ for ISM
fluids is due to difficulties [15, 16] when constructing the transverse
part of $\bms{\hat P}$. While the longitudinal part $\bms{\hat P}_{\rm L}$
can easily be expressed in terms of the operator of charge density, to
define the transverse component $\bms{\hat P}_{\rm T}$ involving additional
dynamical variables is necessary. There are two approaches for describing
the electromagnetic phenomena in ISMs. They differ between themselves in
the way of how to treat the current of moving charges. In the abbreviated
approach [17], electric and magnetic parts of the total current are
undistinguishable from one another in the presence of spatially
inhomogeneous fields, i.e, when $k \ne 0$. In such a case, the magnetic
part is included into a generalized $\bms{\hat P}$-vector and the
electromagnetic phenomena in the system are determined by a transverse
dielectric permittivity $\varepsilon \scs{\rm T}(k,\omega)$ which is directly
connected with the generalized conductivity $\sigma \scs{\rm T}(k,\omega)=
\frac{{\rm i} \omega}{4\pi} (\varepsilon \scs{\rm T}(k,\omega)-1)$. The
second approach [18] is based on a separation of the microscopic current
into electric and magnetic parts at arbitrary wavenumbers. This leads to
the introduction of two functions, $\epsilon \scs{\rm T}(k,\omega) \ne
\varepsilon \scs{\rm T}(k,\omega)$ and $\mu(k,\omega)$, describing the
transverse fluctuations of polarization and magnetization densities
separately. The last function is associated with a generalized magnetic
permittivity of ISM systems. In this approach the transverse dielectric
permittivity $\varepsilon \scs{\rm T}(k,\omega)$ can easily be reproduced
using functions $\epsilon \scs{\rm T}(k,\omega)$ and $\mu(k,\omega)$. In
the infinite-wavelength limit $k \to 0$, when the spatial dispersion can
be neglected, the two approaches become completely equivalent. In this case
the electric phenomena are uniquely described by the frequency-dependent
dielectric constant $\varepsilon(\omega)=\lim_{k \to 0} \epsilon \scs{\rm
T}(k,\omega)=\lim_{k \to 0} \varepsilon \scs{\rm L,T}(k,\omega)$, whereas a
magnetic state is determined by the magnetic permittivity $\mu(\omega)=
\lim_{k\to0} \mu(k,\omega)=[1-\frac{\omega^2}{c^2} \lim_{k\to0} (\varepsilon
\scs{\rm T}(k,\omega)-\varepsilon \scs{\rm L}(k,\omega))/k^2]^{-1}$, where
$c$ denotes the velocity of light [17].

Until recently, quite a few papers [19, 20] dealt with the investigation of
correlations related to the current of charges in ISMs. In these articles,
spectra of the longitudinal and transverse components of the hydrogen
current, on the time scale peculiar to the librational dynamics of water,
were calculated for the TIP4P potential at small wavenumbers. The first
calculation of the transverse dielectric function $\epsilon \scs{\rm T}(k)=
\lim_{\omega \to 0} \epsilon \scs{\rm T}(k,\omega)$ has been performed
by Raineri and Friedman [16], but in the static regime only and for the
simplest $\xi$DS model. At the same time, there were no attempts to
investigate the entire wavevector and frequency dependence of the
transverse dielectric function and the magnetic susceptibility
for ISM fluids.

In the present paper the dielectric properties of ISMs are investigated on
the basis of current correlations, including their separation into electric
and magnetic parts. This allows us to determine both the longitudinal and
transverse components of the dielectric permittivity $\varepsilon(k,\omega)$
as well as the dielectric $\epsilon \scs{\rm T}(k,\omega)$ and magnetic
$\mu(k,\omega)$ functions. Actual MD simulations are performed for the TIP4P
model of water using the interaction site reaction field (ISRF) geometry [3]
and the Ewald method. The results obtained for $\epsilon \scs{\rm T}(k,
\omega)$ and $\mu(k,\omega)$ are presented in a wide region of wavevectors
and frequencies.

\section{Electromagnetic fluctuation formulas for ISM fluids}

\subsection{Generalized dielectric constant and conductivity}

We shall consider a polar fluid consisting of $N$ identical molecules which
are composed of $M$ interaction sites and enclosed in a volume $V$. The
microscopic density of charges for such a system at point $\bms{r} \in V$
and time $t$ is of the form \mbox{$\hat Q(\bms{r},t) = \sum_{i=1}^N
\sum_{a=1}^M q \scs{a} \delta(\bms{r}-\bms{r}_i^a(t))$}, where $q \scs{a}$
and $\bms{r}_i^a(t)$ are the charge and position of site $a$ within molecule
$i$, respectively. Shifting from the real coordinate space $\{\bms{r}\}$
into the $\{\bms{k}\}$-representation by the spatial Fourier transform
$\{\bms{k}\}=\displaystyle \int_V \{\bms{r}\} {\mbox{\large e}}^{-{\rm i}
\bvs{k\!\cdot\!r}} {\rm d} \bms{r}$, we obtain that $\hat Q(\bms{k},t) =
\sum_{i=1}^N \sum_{a=1}^M q \scs{a} {\mbox{\large e}}^{-{\rm i} \bvs{k\!
\cdot\!r}_i^a(t)}$. For the investigation of dielectric properties, it is
convenient to introduce the microscopic vector $\bms{\hat P}(\bms{k},t)$
of polarization density. The longitudinal part $\bms{\hat P}_{\rm L} =
\bms{\hat k} \, \bms{\hat P} \bms{\cdot} \bms{\hat k}$ of this vector can
be determined using the relation $\hat Q(\bms{r},t)=-{\rm div} \bms{\hat P}
(\bms{r},t)$ in $\bms{k}$-space, i.e., $\hat Q(\bms{k},t)=-{\rm i} \bms{k}
\bms{\cdot} \bms{\hat P}(\bms{k},t)$. Then we find
\begin{equation}
\bms{\hat P}_{\rm L}(\bms{k},t) = \frac{{\rm i}\bms{k}}{k^2}
\hat Q(\bms{k},t) = \frac{{\rm i}\bms{\hat k}}{k} \sum_{i=1}^N \sum_{a=1}^M
q \scs{a} {\mbox{\large e}}^{-{\rm i} \bvs{k\!\cdot\!r}_i^a(t)} \ ,
\end{equation}
where $\bms{\hat k}=\bms{k}/k$ is the unit vector directed along $\bms{k}$.

Recently [1], it has been shown that the longitudinal component $\varepsilon
\scs{\rm L}(k,\omega)$ of the wavevector- and frequency-dependent dielectric
tensor can be calculated in computer experiment via the fluctuation formula
\begin{equation}
\frac{\varepsilon \scs{\rm L}(k,\omega) - 1}{\varepsilon
\scs{\rm L}(k,\omega)}=\frac{h {\scr L}_{{\rm i}\omega}
(-\dot{G}_{\rm L}(k,t))}{1+h D(k)
{\scr L}_{{\rm i}\omega} (-\dot{G}_{\rm L}(k,t))} =
h {\scr L}_{{\rm i}\omega} (-\dot{g} \scs{\rm L}(k,t)) \ .
\end{equation}
Here, $G_{\rm L}(k,t)=\left<\bms{\hat P}_{\rm L}(\bms{k},0) \bms{\cdot}
\bms{\hat P}_{\rm L}(-\bms{k},t) \right> \scs{0} \Big/ N d^2$ is the
longitudinal wavevector-dependent dynamical Kirkwood factor computed in
simulation for a finite sample, $\left< \ \ \right> \scs{0}$ denotes the
statistical averaging in the absence of external fields, $d$ designates
the permanent magnitude of molecule's dipole moment $\bms{d}_i=\sum_a^M
q \scs{a} \bms{r}_i^a$, the Laplace transform is defined as ${\scr L}_
{{\rm i}\omega}(\{t\}) = \displaystyle \int_{0}^{\infty} \{t\} \ {\mbox
{\large e}}^{-{\rm i}\omega t} {\rm d} t$, \, $k_{\rm B}$ and $T$ are the
Boltzmann's constant and the temperature of the system, respectively,
$h=4\pi N d^2 \Big/ Vk_{\rm B}T$ and $\dot{G}_{\rm L}(k,t) \equiv \partial
G_{\rm L}(k,t)/\partial t$. The function $D(k)$ takes into account boundary
conditions, applied in simulation, and for finite samples it always is equal
to 1 at $k=0$. For nonzero wavevectors this function tends to zero in the
case of Ewald summation, while $D(k)=3 j\scs{1}(kR)/(kR)$ within the
reaction field geometry [1, 3], where $R$ and $j\scs{1}(z)=\sin(z)/z^2-
\cos(z)/z$ are the cut-off radius and spherical Bessel function of first
order, respectively. For infinite systems ($R \to \infty$) the function
$D(k)$ is equal to zero at arbitrary wavenumbers and the computer adapted
formula (2) reduces to the well-known fluctuation formula in terms of the
infinite-system Kirkwood factor $g \scs{\rm L}(k,t)=\lim_{N \to \infty}
G_{\rm L}(k,t)$.

As we can see, using the microscopic operator $\hat Q(\bms{k},t)$ of charge
density allows one to determine uniquely only the longitudinal part of
polarization vector $\bms{\hat P}(\bms{k},t)$. To construct the transverse
part, it is necessary to involve additional dynamical variables. We shall
show now how to derive fluctuation formulas which give the possibility to
computer both the longitudinal as well as the transverse component of
the dielectric constant of ISMs.

The processes of dynamical polarization in dielectrics cause a current of
charges. This current is described by the microscopic operator of current
density $\bms{\hat I}$ and may be associated with the generalized
polarization current [17], so that
\begin{equation}
\bms{\hat I}(\bms{k},t)=\sum_{i, a}^{N,M} q \scs{a} \bms{V}_i^a(t)
{\mbox{\large e}}^{-{\rm i} \bvs{k\!\cdot\!r}_i^a(t)} =
\frac{\rm d}{{\rm d} t} \bms{\hat P}(\bms{k},t) \ ,
\end{equation}
where $\bms{V}_i^a(t)$ denotes the velocity of site $a$ within the molecule
$i$ at time $t$. The relation (3) can be considered as a more general
definition of $\bms{\hat P}$-vector, because it allows to determine its
longitudinal $\bms{\hat P}_{\rm L}=\bms{\hat k} \, \bms{\hat P} \bms{\cdot}
\bms{\hat k}$ and transverse $\bms{\hat P}_{\rm T}=(\bms{1}-\bms{\hat k}
\bms{\hat k}) \bms{\hat P}=\bms{\hat k} \bms{\times} [\bms{\hat P} \bms
{\times} \bms{\hat k}]$ components, where $\bms{1}$ is the unit tensor of
the second rank. It is worth to emphasize that the generalized vector
$\bms{\hat P}$ of polarization induction includes both electric as well
as magnetic contributions (see a more detailed discussion of this point
in the next subsection). In a particular case of longitudinal polarization
this vector satisfies completely our previous definition (1) of $\bms{\hat
P}_{\rm L}$. Indeed, taking the full derivative of (1) with respect to time,
it can be verified easily on the basis of equation (3) that $\frac{\rm d}
{{\rm d} t} \bms{\hat P}_{\rm L}(\bms{k},t)=\hat{\bms{k}} \, \bms{\hat k}
\bms{\cdot} \bms{\hat I}(\bms{k},t) \equiv \bms{\hat I}_{\rm L}(\bms{k},t)$.

Let us apply to the system under consideration an external electric field
$\bms{E}_0(\bms{k},t)$ which contains longitudinal $\bms{E}_0^{\rm L}$ and
transverse $\bms{E}_0^{\rm T}$ components. The total field in the system
$\bms{\hat E}(\bms{k},t)$ consists of the external field and an internal
field of charged sites. Neglecting the relativistic terms due to the
dynamical magnetic field of moving charges, the internal field can be
presented [1, 3] in the purely longitudinal form $-4\pi(1-D(k))\bms{\hat
P}_{\rm L}$. Therefore, the longitudinal and transverse components of the
total field $\bms{\hat E}=\bms{\hat E}_{\rm L}+\bms{\hat E}_{\rm T}$ are
equal to $\bms{\hat E}_{\rm L}(\bms{k},t)=\bms{E}_0^{\rm L}(\bms{k},t)-
4\pi (1-D(k)) \bms{\hat P}_{\rm L}(\bms{k},t)$ and $\bms{\hat E}_{\rm T}
(\bms{k},t)=\bms{E}_0^{\rm T}(\bms{k},t)$, respectively. The longitudinal
$\varepsilon \scs{\rm L}(k,\omega)$ and transverse $\varepsilon \scs{\rm
T}(k,\omega)$ components of the wavevector- and frequency-dependent
dielectric tensor $\bms{\varepsilon}(\bms{k},\omega)=\varepsilon \scs{\rm
T}(k,\omega) \bms{1}+(\varepsilon \scs{\rm L}(k,\omega)-\varepsilon
\scs{\rm T}(k,\omega)) \bms{\hat k} \bms{\hat k}$ can be defined via
the material relations $\bms{P}_{\rm L,T}(\bms{k},\omega)=\frac{1}{4\pi}
(\varepsilon \scs{\rm L,T}(k,\omega) - 1)$ $\bms{E}_{\rm L,T}(\bms{k},
\omega)$, where $\bms{P}_{\rm L,T}(\bms{k},\omega)=\left< \bms{\hat
P}_{\rm L,T}(\bms{k},\omega) \right>$ and $\bms{E}_{\rm L,T}(\bms{k},
\omega)=\left< \bms{\hat E}_{\rm L,T}(\bms{k},\omega)\right>$ are
macroscopic values of the polarization and total field, $\left< \ \
\right>$ denotes the statistical averaging in the presence of the external
field and the time Fourier transform $\{\omega\} = \displaystyle \int_
{-\infty}^{\infty} \{t\} {\mbox {\large e}}^{-{\rm i} \omega t} {\rm d}t$
has been used for functions $\bms{\hat P}_{\rm L,T}(\bms{k},t)$ and
$\bms{\hat E}_{\rm L,T}(\bms{k},t)$.

In the case of dynamical polarization, when $\bms{I}(\bms{k},\omega)=
\left< \bms{\hat I}(\bms{k},\omega) \right> \ne 0$ at $\omega \ne 0$ (note
that static macroscopic currents of coupled charges are equal to zero),
the Maxwell equations can be written in a form as for conductors, namely,
in terms of the generalized polarization conductivity $\sigma(k,\omega)$.
The longitudinal and transverse components $\sigma \scs{\rm L,T}(k,\omega)$
of this wavevector- and frequency-dependent conductivity are defined via
the material relations $\bms{I}_{\rm L,T}(\bms{k},\omega)= \sigma \scs{\rm
L,T}(k,\omega) \bms{E}_{\rm L,T}(\bms{k},\omega)$. In the frequency
representation, equation (3) stays that $\bms{\hat I}_{\rm L,T}(\bms{k},
\omega)={\rm i} \omega \bms{\hat P}_{\rm L,T}(\bms{k},\omega)$. Thus we can
express the dielectric constant in terms of the polarization conductivity
as follows
\begin{equation}
\varepsilon \scs{\rm L,T}(k,\omega)-1=
\frac{4\pi \sigma \scs{\rm L,T}(k,\omega)}{{\rm i} \omega} \ .
\end{equation}
According to the perturbation theory of first order with respect to external
fields, we obtain for the macroscopic current $\bms{I}_{\rm L,T}(\bms{k},
\omega)={\scr L}_{{\rm i} \omega} \left< \bms{\hat I}_{\rm L,T}(\bms{k},0)
\bms{\cdot} \bms{\hat I}_{\rm L,T}(-\bms{k},t) \right> \scs{0} \frac{\bvs
{E}_0^{\rm L,T}(\bvs{k},\omega)}{\{2\} V k_{\rm B} T}$, where the multiplier
\{2\} is to be included for the transverse component only. Then eliminating
the electric fields and using relation (4) yields
\begin{eqnarray}
&&\!\frac{\varepsilon \scs{\rm L}(k,\omega) - 1}
{\varepsilon \scs{\rm L}(k,\omega)}=\frac{h {\scr L}_{{\rm i}\omega}
(C_{\rm L}(k,t))} {{\rm i} \omega + h D(k)
{\scr L}_{{\rm i}\omega} (C_{\rm L}(k,t))} =
\frac{h}{{\rm i}\omega} {\scr L}_{{\rm i}\omega}
(c \scs{\rm L}(k,t)) \ , \\ [12pt]
&& \varepsilon \scs{\rm T}(k,\omega) - 1 =
\frac{h}{{\rm i}\omega} {\scr L}_{{\rm i}\omega}
(c \scs{\rm T}(k,t)) \ ,
\end{eqnarray}
where
\vspace{-6pt}
\begin{equation}
C_{\rm L,T}(k,t)= \frac{\left<\bms{\hat I}_{\rm L,T}(\bms{k},0) \bms{\cdot}
\bms{\hat I}_{\rm L,T}(-\bms{k},t) \right> \scs{0}}{\{2\} N d^2}
\end{equation}
are the longitudinal and transverse components of the wavevector-dependent
dynamical Kirkwood factor of second order for the finite sample, whereas
$c \scs{\rm L,T}(k,t)$ are the corresponding functions of the infinite
system and $c \scs{\rm T}(k,t)=C_{\rm T}(k,t)$.

The fluctuation formulas (5) and (6) can be used in simulations to evaluate
the longitudinal and transverse components of the dielectric constant on the
basis of equilibrium current-current correlations (7). In the case of
longitudinal fluctuations, the formula (5) is mathematically equivalent to
the usual relation (2). Indeed, in view of the equality $\bms{\hat I}_{\rm
L}(\bms{k},t)=\frac{\rm d}{{\rm d} t} \bms{\hat P}_{\rm L}(\bms{k},t)$, it
can be shown easily that $C_{\rm L}(k,t)=-\frac{\partial^2}{\partial t^2}
G_{\rm L}(k,t) \equiv - \ddot {G}_{\rm L}(k,t)$. Then applying the Laplace
transform we obtain ${\scr L}_{{\rm i}\omega}(C_{\rm L}(k,t))=\dot{G}_{\rm
L}(k,0) +{\rm i}\omega {\scr L}_{{\rm i}\omega}(-\dot{G}_{\rm L}(k,t))$,
where $\dot{G}_{\rm L}(k,0)=0$ because the Kirkwood factor $G_{\rm L}(k,t)$
is an even function on time, and we immediately recover the fluctuation
formula (2) from (5). Furthermore, taking into account that ${\scr L}_{{\rm
i}\omega}(-\dot{G}_{\rm L}(k,t)) = G_{\rm L}(k,0) - {\rm i}\omega {\scr L}_
{{\rm i}\omega}(G_{\rm L}(k,t)) \equiv {\scr L}_{{\rm i}\omega}(C_{\rm L}
(k,t))/{\rm i}\omega$ we obtain, in particular, that the static ($t=0$)
longitudinal Kirkwood factor $G_{\rm L}(k) \equiv G_{\rm L}(k,0)$ is
connected with the dynamical Kirkwood factor of second order as
\vspace{-8pt}
\begin{equation}
G_{\rm L}(k)=\lim_{\omega \to +0}
\frac{{\scr L}_{{\rm i}\omega} (C_{\rm L}(k,t))}
{{\rm i} \omega} = - \int_0^\infty t C_{\rm L}(k,t) {\rm d} t \ , \\ [3pt]
\end{equation}
where the equality (8) holds for the infinite-system functions
$g \scs{\rm L}(k)$ and $c \scs{\rm L}(k,t)$ as well.

Despite the difference, in general, between the longitudinal Kirkwood
factors for the finite and infinite systems, we, nevertheless, can
reproduce the infinite-system behaviour indirectly using the relations
\begin{equation}
\frac{1}{{\scr L}_{{\rm i}\omega} (\dot g \scs{\rm L}(k,t))} =
\frac{1}{{\scr L}_{{\rm i}\omega} (\dot G_{\rm L}(k,t))} -h D(k)
\ , \ \ \ \ \
\frac{{\rm i}\omega}{{\scr L}_{{\rm i}\omega} (c \scs{\rm L}(k,t))} =
\frac{{\rm i}\omega}{{\scr L}_{{\rm i}\omega} (C_{\rm L}(k,t))}+h D(k) \ .
\end{equation}
Taking into account the Laplace boundary theorem $\lim_{\omega \to \infty}
{\rm i} \omega {\scr L}_{{\rm i}\omega}(\phi(t)) = \lim_{t \to 0} \phi(t)$,
it can be shown from (9) that $\frac{\partial^2}{\partial t^2} G_{\rm L}(k,t)
\Big|_{t=0}=\frac{\partial^2}{\partial t^2} g \scs{\rm L}(k,t) \Big|_{t=0}$
and $C_{\rm L}(k)=c \scs{\rm L}(k)$, respectively. So that, contrary to the
usual Kirkwood factor $G_{\rm L}(k)$, the static Kirkwood factor of second
order $C_{\rm L}(k)$ is free of boundary effects. The transverse function
$c \scs{\rm T}(k,t)$ is equal to $C_{\rm T}(k,t)$ at arbitrary times,
because $\bms{E}_{\rm T}=\bms{E}_0^{\rm T}$ for nonrelativistic systems,
and it can be obtained in simulations directly without additional
manipulations. Moreover, the static Kirkwood factors of second order
$c \scs{\rm L,T}(k)$ are presented in analytical forms (see Appendix).

\subsection{Transverse dielectric function and magnetic permittivity}

We now consider an alternative approach to describe the electromagnetic
properties of ISM systems. This approach is based on the separation of
macroscopic currents into electric and magnetic parts in the limit
$k,\omega \to 0$. Neglecting terms of order $k^2$, $\omega^2$, $k
\omega$ and higher, the averaged values for the operator of current
density (3) can be evaluated in $(\bms{k},\omega)$-space as
\begin{equation}
\langle \bms{\hat I}(\bms{k},\omega) \rangle = {\rm i} c\bms{k} \bms{\times}
\langle \bms{\hat{\cal M}}(\omega) \rangle + {\rm i} \omega \langle
\bms{\hat{\cal P}}(\omega) \rangle \ ,
\end{equation}
where $\bms{\hat{\cal P}}(\omega)$ and $\bms{\hat{\cal M}}(\omega)$ are
frequency components of the total electric dipole moment $\bms{\hat{\cal
P}}(t)=\sum_{i=1}^N \bms{d}_i(t)$ and the rotational part $\bms{\hat{\cal
M}}(t)=\sum_{i=1}^N \bms{m}_i(t)$ of the magnetic dipole moment of the
system, where $\bms{m}_i=\frac{1}{2c} \sum_{a=1}^M q \scs{a} \bms{\delta}_i^a
\bms{\times} \bms{v}_i^a$. In our notations $\bms{\delta}_i^a=\bms{r}_i^a-
\bms{r}_i$ and $\bms{v}_i^a=\bms{V}_i^a-\bms{V}_i=\bms{\mit \Omega}_i
\bms{\times} \bms{\delta}_i^a$ are the positions and velocities of sites
relatively to the molecular centre of mass $\bms{r}_i$ and its velocity
$\bms{V}_i$, respectively, and $\bms{\mit \Omega}_i$ is the angular
velocity of the $i$th molecule.

Despite the fact that the separation (10) is realized uniquely only in
the macroscopic regime at small wavenumbers and frequencies [17], we shall
apply it at the microscopic level of description and extend to arbitrary
$k$ and $\omega$. Such an extension can be performed by writing (in
$(\bms{k},t)$-representation)
\begin{equation}
\bms{\hat I}(\bms{k},t)={\rm i} c\bms{k} \bms{\times}
\bms{\hat{\cal M}}(\bms{k},t) + \frac{\rm d}{{\rm d} t}
\bms{\hat{\cal P}}(\bms{k},t) \ ,
\end{equation}
where $\bms{\hat{\cal P}}(\bms{k},t)$ and $\bms{\hat{\cal M}}(\bms{k},t)$
are appropriate dynamical variables associated with the microscopic
polarization and magnetization densities, respectively. In view of equations
(3), (10) and (11), these variables should satisfy the limiting transitions
$\lim_{k\to0} \bms{\hat{\cal P}}(\bms{k},t)=\bms{\hat{\cal P}}(t)$ and
$\lim_{k\to0} \bms{\hat{\cal M}}(\bms{k},t)=\bms{\hat{\cal M}}(t)$. Moreover,
taking the scalar product of equations (3) and (11) on unit vector $\bms{\hat
k}$ yields the condition $\bms{\hat{\cal P}}_{\rm L}(\bms{k},t)=\bms{\hat
P}_{\rm L}(\bms{k},t)$, where the equality $\bms{\hat k} \bms{\cdot}
[\bms{k} \bms{\times} \bms{\hat{\cal M}}]=0$ has been used.

The explicit expression for $\bms{\hat{\cal P}}(\bms{k},t)$ can be derived
using the following procedure [16]. Let us split the operator $\hat Q(\bms{k},
t)=\sum_{i=1}^N \hat q \scs{i}(\bms{k},t) {\mbox {\large e}}^{-{\rm i} \bvs
{k\!\cdot\!r}_i(t)}$ of charge density into its molecular components $\hat
q \scs{i}(\bms{k},t)=\sum_{a=1}^M q \scs{a} {\mbox {\large e}}^{-{\rm i}
\bvs{k\!\cdot\!\delta}_i^a(t)}$. Then it is natural to introduce the total
microscopic polarization density as $\bms{\hat {\cal P}} (\bms{k},t)=
\sum_{i=1}^N \bms{\hat p}_i(\bms{k},t) {\mbox {\large e}}^{-{\rm i} \bvs{k\!
\cdot\!r}_i(t)}$, where $\bms{\hat p}_i(\bms{k},t)$ is the polarization
density of the $i$th molecule, and to extend the relation $\hat Q(\bms{k},t)=
-{\rm i} \bms{k}\bms{\cdot} \bms{\hat P}(\bms{k},t)$ to the molecular level,
i.e., $\hat q \scs{i}(\bms{k},t)=-{\rm i} \bms{k} \bms{\cdot} \bms{\hat p}_i
(\bms{k},t)$. Further, taking into account the identity ${\mbox {\large
e}}^{\xi}=1+\xi \int_0^1 {\rm d} u {\mbox {\large e}}^{u\xi}$, applied to the
quantity $\xi=-{\rm i} \bms{k\!\cdot\!\delta}_i^a$, and using the molecular
charge electroneutrality $\sum_{a=1}^M q \scs{a}=0$, we present $q \scs{i}
(\bms{k},t)$ as $-{\rm i} \bms{k} \bms{\cdot} \sum_{a=1}^M q \scs{a} \bms
{\delta}_i^a(t) \int_0^1 {\rm d} u {\mbox {\large e}}^{-{\rm i} u \bvs{k\!
\cdot\!\delta}_i^a(t)}$. The last expression leads to $\bms{\hat p}_i(\bms
{k},t)=\sum_{a=1}^M q \scs{a} \bms{\delta}_i^a(t) \int_0^1 {\rm d} u {\mbox
{\large e}}^{-{\rm i} u \bvs{k\!\cdot\!\delta}_i^a(t)}$ and, therefore,
\begin{equation}
\bms{\hat {\cal P}}(\bms{k},t)=\sum_{i=1}^N \sum_{a=1}^M q \scs{a}
\bms{\delta}_i^a(t) \frac{
{\mbox {\large e}}^{-{\rm i} \bvs{k\!\cdot\!\delta}_i^a(t)}-1}
{-{\rm i} \bms{k\!\cdot\!\delta}_i^a(t)}\,
{\mbox {\large e}}^{-{\rm i} \bvs{k\!\cdot\!r}_i(t)} \ .
\end{equation}
Of course, such a procedure does not define $\bms{\hat {\cal P}}(\bms{k},t)$
uniquely, because $q \scs{i}$ is indifferent to the transverse part of
$\bms{\hat p}_i$. Nevertheless, it was assumed [16] to adopt equation (12)
as the definition of microscopic polarization density for ISM fluids. From
this definition it immediately follows that the longitudinal components of
vectors $\bms{\hat P}$ and $\bms{\hat{\cal P}}$ coincide completely
between themselves at arbitrary wavevectors.

Since the functions $\bms{\hat I}(\bms{k},t)$ and $\bms{\hat {\cal P}}
(\bms{k},t)$ are already defined, equation (11) allows one to determine the
transverse part of microscopic magnetization density,
\begin{equation}
\bms{\hat{\cal M}}_{\rm T}(\bms{k},t)=
\bms{\hat k} \bms{\times} [\bms{\hat{\cal M}}(\bms{k},t)
\bms{\times} \bms{\hat k}] = \frac{1}{{\rm i} ck}
\Big (\bms{\hat I}(\bms{k},t)-\frac{\rm d}{{\rm d}t}
\bms{\hat {\cal P}}(\bms{k},t) \Big) \bms{\times} \bms{\hat k} \ .
\end{equation}
It can be verified using equations (3), (12) and (13) that in the
infinite-wavelength limit $k\to0$ the functions $\bms{\hat{\cal P}}
(\bms{k},t)$ and $\bms{\hat{\cal M}}_{\rm T}(\bms{k},t)$ tend to the
true microscopic variables $\sum_{i=1}^N \bms{d}_i(t)=\sum_{i=1}^N
\sum_{a=1}^M q \scs{a} \bms{\delta}_i^a(t)$ and $\bms{\hat k} \bms{\times}
\sum_{i=1}^N [\bms{m}_i(t) \bms{\times} \bms{\hat k}]$ which correspond
to the electric and rotational magnetic dipole moments of the system,
respectively. However, at $k \ne 0$ the vectors $\bms{\hat{\cal P}}(\bms
{k},t)$ (as well as $\bms{\hat P}(\bms{k},t)$) and $\bms{\hat{\cal M}}_{\rm
T}(\bms{k},t)$ take into account explicitly the charge distribution within
a finite spatial extend of the molecule and, therefore, they can not longer
be associated with the point dipole densities $\hat{\scr P}(\bms{k},t)=
\sum_{i=1}^N \bms{d}_i(t) {\mbox {\large e}}^{-{\rm i} \bvs{k\!\cdot\!r}_i
(t)}$ and ${\hat{\scr M}}_{\rm T}(\bms{k},t)=\bms{\hat k} \bms{\times}
\sum_{i=1}^N [\bms{m}_i(t) \bms{\times} \bms{\hat k}] {\mbox {\large e}}^{-
{\rm i} \bvs{k\!\cdot\!r}_i(t)}$. For purely dipole models, when $|q \scs{a}|
\to \infty$ and $\max_a|\bms{\delta}_i^a| \to 0$, provided $d \to$ const,
we obtain $\bms{\hat{\cal P}}(\bms{k},t) \to \hat{\scr P}(\bms{k},t)$, but
${\hat{\scr M}}(\bms{k},t) \to 0$ because of $\bms{m}_i \to 0$ and, thus,
there is no magnetic response in this case, i.e., $\bms{\hat{\cal P}}(\bms
{k},t) \equiv \bms{\hat P}(\bms{k},t)$ at arbitrary wavevectors.

In the present approach the electric phenomena are described by the
frequency-dependent dielectric functions $\epsilon \scs{\rm L,T}(k,\omega)$,
while the magnetic state is determined by the magnetic permittivity $\mu
(k,\omega) \equiv \mu \scs{\rm T}(k,\omega)$ (for the system under
consideration the longitudinal magnetic susceptibility is absent).
These quantities are defined via the material relations $\bms{\cal P}_{\rm
L,T}(\bms{k},\omega)=\frac{1}{4\pi} (\epsilon \scs{\rm L,T}(k,\omega) - 1)
\bms{E}_{\rm L,T}(\bms{k},\omega)$ and $\bms{\cal M}_{\rm T}(\bms{k},
\omega)=\frac{1}{4\pi} (\mu(k,\omega) - 1) \bms{H}_{\rm T}(\bms{k},\omega)$,
where the statistical averaging of $\bms{\cal P}=\langle \bms{\hat{\cal P}}
\rangle$ and $\bms{{\cal M}}=\langle \bms{\hat{\cal M}}\rangle$ is performed
in the presence of external electric $\bms{E}_0$ and magnetic $\bms{H}_0$
fields. We mention that for nonrelativistic systems the total magnetic field
$\bms{H}$ is indistinguishable from the external field, i.e., $\bms{H}_{\rm
T}=\bms{H}_0^{\rm T}$. We stress also that it is necessary to distinguish
the generalized dielectric tensor $\varepsilon \scs{\rm L,T}(k,\omega)$
corresponding to the total current $\bms{I}={\rm d} \bms{P}/{\rm d} t$ from
the dielectric functions $\epsilon \scs{\rm L,T}(k,\omega)$ related to the
electric part ${\rm d} \bms{\cal P}/{\rm d} t$ of $\bms{I}$. It is obvious
that $\epsilon \scs{\rm L}(k,\omega) \equiv \varepsilon \scs{\rm L}(k,
\omega)$ because of $\bms{\hat {\cal P}}_{\rm L}(\bms{k},t)=\bms{\hat
P}_{\rm L}(\bms{k},t)$, but $\epsilon \scs{\rm T}(k,\omega) \ne
\varepsilon \scs{\rm T}(k,\omega)$ since, in general, $\bms{\hat {\cal
P}}_{\rm T}(\bms{k},t) \ne \bms{\hat P}_{\rm T}(\bms{k},t)$. The
fluctuation formula for the transverse dielectric function $\epsilon
\scs{\rm T}(k,\omega)$ can be obtained in a similar way as for
permittivities $\varepsilon \scs{\rm L,T}(k,\omega)$ (see the preceding
subsection and Ref. [1]). The result is
\begin{equation}
\epsilon \scs{\rm T}(k,\omega) - 1 =
h {\scr L}_{{\rm i}\omega} (-\dot{g} \scs{\rm T}(k,t)) \ ,
\end{equation}
where the infinite-system correlation function $g \scs{\rm T}(k,t) =
\left<\bms{\hat{\cal P}}_{\rm T}(\bms{k},0) \bms{\cdot} \bms{\hat{\cal
P}}_{\rm T}(-\bms{k},t) \right> \scs{0} \Big/ 2 N d^2$ can be evaluated
in simulation directly, because of $\bms{E}_{\rm T}=\bms{E}_0^{\rm T}$.

Finally, we would like to discuss about a relationship of the two
approaches presented in more detail. In the abbreviated approach the
electric and magnetic parts in the total current are not distinguished
from one another. In such a situation the magnetic part is included into
the generalized $\bms{\hat P}$-vector (3) and the transverse electromagnetic
fluctuations are described by one function only, namely, by the generalized
dielectric constant $\varepsilon \scs{\rm T}(k,\omega)$. This including
indeed can be realized because at $k, \omega \ne 0$ magnetic fields
are expressed in terms of electric fields $\bms{E}$ using the Maxwell
equation $c\,{\rm rot} \bms{E} = - \partial \bms{B} /\partial t$ in
$(\bms{k},\omega)$-representation, i.e., $c \bms{k} \bms{\times} \bms{E}
({\bms{k},\omega}) = - \omega \bms{B}_{\rm T}(\bms{k},\omega)$, where
$\bms{B}_{\rm T}(\bms{k},\omega)=\mu(k,\omega) \bms{H}_{\rm T}(\bms{k},
\omega)$. At the same time, the extended approach involves two quantities,
$\epsilon \scs{\rm T}(k,\omega)$ and $\mu(k,\omega)$, for the description.
As far as these two approaches deal with the same microscopic current, the
quantities $\varepsilon \scs{\rm T}(k,\omega)$, $\epsilon \scs{\rm T}(k,
\omega)$ and $\mu(k,\omega)$ are not independent. The corresponding relation
can be found comparing the right-hand sides of equations (3) and (11)
between themselves and using material equations. Then one obtains
\begin{equation}
\varepsilon \scs{\rm T}(k,\omega) = \epsilon \scs{\rm T}(k,\omega) +
\frac{c^2 k^2}{\omega^2} \frac{\mu(k,\omega)-1}{\mu(k,\omega)}
\end{equation}
and, therefore, the wavevector- and frequency-dependent magnetic
permittivity is caused by the difference between functions $\varepsilon
\scs{\rm T}(k,\omega)$ and $\epsilon \scs{\rm T}(k,\omega)$.

The simplest way to obtain the explicit fluctuation formula for the magnetic
permittivity is based on relation (15) and fluctuation formulas (6) and (14)
for $\varepsilon \scs{\rm T}(k,\omega)$ and $\epsilon \scs{\rm T}(k,\omega)$.
In view of equation (11), the transverse current autocorrelation function $c
\scs{\rm T}(k,t)$ (7), appearing in fluctuation formula (6) for $\varepsilon
\scs{\rm T}(k,\omega)$, can be expressed in terms of the polarization $g
\scs{\rm T}(k,t)$ and magnetization $s \scs{\rm T}(k,t) = \left<\bms{\hat
{\cal M}}_{\rm T}(\bms{k},0) \bms{\cdot} \bms{\hat{\cal M}}_{\rm T}(-\bms
{k},t) \right> \scs{0} \Big/ 2 N d^2$ correlation functions as $c \scs{\rm
T}(k,t)=c^2 k^2 s \scs{\rm T}(k,t)-\partial^2 g \scs{\rm T}(k,t)/\partial
t^2$. It is necessary to point out that for spatially homogeneous systems,
as in our case, the cross function $\left<\bms{\hat{\cal P}}_{\rm T} \bms
{\cdot} \bms{\hat{\cal M}}_{\rm T} \right> \scs{0}$, corresponding to
correlations between the polarization and magnetization vectors, does not
appear and it is equal to zero for arbitrary wavenumbers and times. Indeed,
let us direct $\bms{\hat k}$-vector along $z$-axis of the laboratory
reference frame. Then it follows from the structure of equation (11) that
$\bms{\hat{\cal P}}_{\rm T} \bms{\cdot} \bms{\hat{\cal M}}_{\rm T} = \frac12
(\hat{\cal P}_x \hat{\cal M}_y + \hat{\cal P}_y \hat{\cal M}_x)$, where
$\hat{\cal P}_{x,y}$ and $\hat{\cal M}_{x,y}$ are the $x$th and $y$th
components of vectors $\bms{\hat{\cal P}}$ and $\bms{\hat{\cal M}}$,
respectively. Since the fluctuations of vector quantities in different
directions of the fixed laboratory frame are statistically independent at
equilibrium, we have that $\left<\bms{\hat{\cal P}}_{\rm T}(\bms{k},0)
\bms{\cdot} \bms{\hat{\cal M}}_{\rm T}(-\bms{k},t) \right> \scs{0}=0$.
Further, as far as the relativistic effects have been neglected, the
functions $\varepsilon \scs{\rm T}(k,\omega)$ and $\epsilon \scs{\rm T}
(k,\omega)$ are evaluated via fluctuation formulas (6) and (14), in fact,
with a precision of order $(v/c)^2$, where $v \ll c$ is the mean heat
velocity of atoms. This leads to uncertainties of order $c^{-4}$ in the
evaluation of $\mu(k,\omega)$ via relation (15). Within the same precision
we can putt $(\mu(k,\omega)-1)/\mu(k,\omega) \approx \mu(k,\omega)-1$, so
that the desired fluctuation formula is
\begin{equation}
\mu(k,\omega)-1=-{\rm i} \omega h {\scr L}_{{\rm i}\omega}
(s \scs{\rm T}(k,t)) + {\cal O}(c^{-4}) \ ,
\end{equation}
where the infinite-system function $s \scs{\rm T}(k,t)$ can be reproduced
directly in simulation, because of $\bms{H}_{\rm T}=\bms{H}_0^{\rm T}$.

From the afore said, it is obvious that the two approaches are completely
equivalent for the evaluations of the generalized dielectric permittivity
$\varepsilon \scs{\rm T}(k,\omega)$ and they can be applied with equal
successes to theoretical applications. However, the extended description
is only the way to determine the magnetic permittivity at $k\ne0$. It is
interesting to point out also that in the infinite-wavelength regime $k
\to0$, the functions $\varepsilon \scs{\rm L}(k,\omega) \equiv \epsilon
\scs{\rm L}(k,\omega)$ and $\epsilon \scs{\rm T}(k,\omega)$ differ between
themselves by terms of order $k^4$ and higher, i.e., $\lim_{k\to0}
[\varepsilon \scs{\rm L}(k,\omega)-\epsilon \scs{\rm T}(k,\omega)]/
k^2=0$. This statement can be examined on the basis of fluctuation
formulas (2), (14) and explicit expression (12) for microscopic variable
$\bms{\hat{\cal P}}$. Then, using relation (15), we obtain that in the
abbreviated description the frequency-dependent magnetic permittivity
$\mu(\omega) = \lim_{k\to0} \mu(k,\omega)$ appears as a result of the
limiting transition
\vspace{-7pt}
\begin{equation}
1 - \frac{1}{\mu(\omega)} = \frac{\omega^2}{c^2}
\lim_{k \to 0} \frac{\varepsilon \scs{\rm T}(k,\omega)-
\varepsilon \scs{\rm L}(k,\omega)}{k^2} \\ [7pt]
\end{equation}
which shows that differences between the longitudinal and transverse
components of the generalized dielectric tensor $\bms{\varepsilon}(\bms{k},
\omega)$ in the infinite-wavelength limit are caused by magnetic properties
of the system [17]. Thus, in the presence of spatially inhomogeneous fields,
the electromagnetic state is determined by the generalized transverse
$\varepsilon \scs{\rm T}(k,\omega)$ and longitudinal $\varepsilon \scs{\rm
L}(k,\omega)$ dielectric functions. When the spatial dispersion can be
neglected, all electric and magnetic phenomena in the system are uniquely
described by two quantities again, namely, by the frequency-dependent
dielectric constant $\varepsilon(\omega)=\lim_{k \to 0} \varepsilon \scs
{\rm L}(k,\omega)=\lim_{k \to 0} \epsilon \scs{\rm L,T}(k,\omega)$ and
magnetic permittivity $\mu(\omega)$.

\section{Numerical calculations for the TIP4P model}

Molecular dynamics simulations were carried out for the TIP4P potential [21]
at a density of 1 g/cm$^3$ and at a temperature of $T=293$ K. We performed
two runs corresponding to the ISRF [3] and Ewald [22] geometries. The
equations of motion were integrated on the basis of a matrix method [23]
with a time step of 2 fs. The observation times over the equilibrium state
were 500 000 and 1 000 000 time steps for the Ewald and ISRF geometry,
respectively. The time correlation functions were calculated in an interval
of 2 ps for the wavenumbers $k=[0,1,\ldots,300] k_{\rm min}$, where $k_{\rm
min}=2 \pi/L=0.319{\rm \AA}^{-1}$ and $L$ is the length of the simulation
box edge. The parameters $\eta =5.76/L$ and $k_{\rm max} = 5 k_{\rm min}$
have been used in the Ewald summation of the Coulomb forces. Other details
of the simulations are similar to those reported earlier [1].

\subsection{Dielectric properties}

The longitudinal component $G_{\rm L}(k)$ of the static wavevector-dependent
Kirkwood factor, calculated in the simulations within the Ewald and ISRF
geometries, is shown in fig.~1a by the full squares and dashed curve,
respectively. Since, in the ISRF geometry the function $D(k)=3 j \scs{1}
(kR)/(kR)$ differs from zero considerably, especially at small wavenumbers,
to evaluate the infinite-system Kirkwood factor $g \scs{\rm L}(k)$ the
performance of self-consistent transformations (9) is necessary. This result
is plotted by the solid curve. Within the Ewald geometry [22] the function
$D(k)$ is equal to $1-\int_0^R k j\scs{1}(k\rho) \Big( {\rm erfc}(\eta \rho)
+\frac{2 \eta}{\sqrt{\pi}} \rho \exp(-\eta^2 \rho^2) \Big) {\rm d} \rho -
\exp(-k^2/4\eta^2)$, where the last term is to be included only if $0 < k
\le k_{\rm max}$, and at the given parameters of the summation $D(k)$ is
very close to zero (${\rm max}_{k\ne0} |D(k)| < 0.00005$). So that the
infinite-system Kirkwood factor is equivalent to that, obtained in the
simulations (excepting the case $k=0$). As we can see from the figure, the
both Ewald and ISRF methods lead to identical results for $g \scs{\rm L}(k)$.
The static Kirkwood factor of second order, $C_{\rm L,T}(k)$ (equation (7)),
is presented in fig.~1b. As was pointed out earlier, this function is free of
boundary conditions and the infinite-system dependence $c \scs{\rm L,T}(k)$
can be reproduced directly in the simulations. This is confirmed by our
calculations performed in the different geometries. The obtained values for
$C_{\rm L,T}(k)$ are practically indistinguishable from those evaluated from
analytical expressions (A5) for $c \scs{\rm L,T}(k)$. They differ from one
another within statistical noise only.

Samples of the normalized, dynamical Kirkwood factor of second order,
$\Psi_{\rm L,T}(k,t)=C_{\rm L,T}(k,t)/C_{\rm L,T}(k)$, are plotted in
fig.~2. The longitudinal infinite-system functions $c \scs{\rm L}(k,t)/c
\scs{\rm L}(k)$ within the ISRF geometry at arbitrary wavenumbers and for
the Ewald geometry at $k=0$ have been determined applying the inverse
Laplace transform to relations (9), whereas $c \scs{\rm T}(k,t) = C_{\rm T}
(k,t)$ for the transverse functions. We note that $C_{\rm L}(k,t)=C_{\rm T}
(k,t)$ at $k=0$. At $k \ne 0$ the Ewald method reproduces directly the
infinite-system behaviour. The agreement between the two sets of data for
$c \scs{\rm L,T}(k,t)$, corresponding to the ISRF and Ewald geometries,
is quite good. It is worth to remark also that the zeroth time moment
$\int_0^\infty c \scs{\rm L}(k,t) {\rm d} t$ is equal to zero for arbitrary
wavenumbers (this statement directly follows from equality (8)), whereas,
in general, $\int_0^\infty c \scs{\rm T}(k,t) {\rm d} t \ne 0$ in the case
of transverse total current fluctuations.

As was shown in the preceding section, the longitudinal static Kirkwood
factor $G_{\rm L}(k)$ of order zero can be determined through the first
time moment on the dynamical Kirkwood factor $C_{\rm L}(k,t)$ of second
order (see equation (8)). The calculated in such a way function $G_{\rm L}
(k)$ within the ISRF geometry and the function $g \scs{\rm L}(k)$ (via $c
\scs{\rm L}(k,t))$ using the Ewald method are presented in fig.~1a by
rotated and direct crosses, respectively. As we see from the calculations,
these functions differ considerably from those obtained in the usual way.
This difference occurs because the calculation of the expression ${\scr
L}_{{\rm i}\omega} (C_{\rm L}(k,t))/{\rm i} \omega$ at small frequencies
is very sensitive to the precision of the evaluation of ${\scr L}_{{\rm i}
\omega} (C_{\rm L}(k,t))$ (dividing two small quantities between themselves).
The problem is complicated additionally because this evaluation requires a
numerical integration of time correlation functions which are defined in
simulations approximately within a statistical noise. Therefore, this method
is not recommended for the investigation of the longitudinal dielectric
constant at low frequency values. However, in the opposite infinite-frequency
limit, the calculation via equation (5) can be more convenient than using
the usual way (2). For example, in this limit, where $\varepsilon \scs{\rm
L}(k,\omega) \to 1$, we obtain from (2) that ${\scr L}_{{\rm i}\omega}
(-\dot{g} \scs{\rm L}(k,t)) = g \scs{\rm L}(k) - {\rm i}\omega {\scr
L}_{{\rm i}\omega} (g \scs{\rm L}(k,t)) \to 0$. The exact computation of
${\rm i}\omega {\scr L}_{{\rm i}\omega} (g \scs{\rm L}(k,t))$ at large
frequencies may lead to a problem (multiplying of quantities with
significantly different orders). This situation is absent in the
case when fluctuation formula (5) is applied to calculations.

And now we are in a position to discuss a behaviour of the generalized
dielectric permittivity in the low frequency regime $\omega \to 0$. As was
established previously [1], the real part of the longitudinal dielectric
permittivity $\varepsilon \scs{\rm L}(k,\omega)$ at $\omega \to 0$ tends to
its static value $\varepsilon \scs{\rm L}(k)$, while the imaginary part
vanishes at arbitrary wavenumbers except the cases of two singularities,
where $\varepsilon \scs{\rm L}(k) = \pm \infty$. In the singularity range,
the real part of the dielectric permittivity is equal to zero, whereas the
imaginary part behaves as $-{\rm i}/\omega \tau_{\rm L}^{\rm cor}(k)$, where
$\tau_{\rm L}^{\rm cor}(k)=\int_0^\infty {\rm d} t\,g \scs{\rm L}(k,t)/g
\scs{\rm L}(k)$ is the longitudinal correlation time. Such a behaviour of
the longitudinal dielectric constant causes the coefficient $\sigma \scs{\rm
L}(k)=\lim_{\omega \to 0} \sigma \scs{\rm L}(k,\omega)=1/4\pi \tau_{\rm
L}^{\rm cor}(k) \ne 0$ for longitudinal conductivity in the singularity
regions, whereas outside the singularities $\sigma \scs{\rm L}(k)=0$. The
existence of a nonvanishing coefficient in the static limit does not lead,
however, to macroscopic currents (as it must be for dielectrics), because
the total longitudinal field in the system vanishes when $|\varepsilon
\scs{\rm L}(k)| \to \infty$.

In the case of transverse fluctuations the pattern is qualitatively
different. According to fluctuation formula (6), the transverse
dielectric permittivity $\varepsilon \scs{\rm T}(k,\omega)$ at low
frequencies behaves as $1 + \lim_{\omega \to 0} h {\scr L}_{{\rm i}
\omega}(c \scs{\rm T}(k,t))/{\rm i}\omega = \varepsilon'(k) - {\rm i}
\varepsilon''(k)$, where $\varepsilon'(k)=\Re \lim_{\omega \to 0}
\varepsilon \scs{\rm T}(k,\omega)=1-h \int_0^\infty t c \scs{\rm T}(k,t)
{\rm d} t$ denotes the real part, whereas the imaginary part $\varepsilon''
(k)=-\Im \lim_{\omega \to 0} \varepsilon \scs{\rm T}(k,\omega)=4 \pi
\sigma \scs{\rm T}(k)/\omega$ is described by the wavevector-dependent
generalized transverse conductivity $\sigma \scs{\rm T}(k)=\frac{h}{4\pi}
\lim_{\omega \to 0}{\scr L}_{{\rm i}\omega}(c \scs{\rm T}(k,t))=\frac{h}
{4\pi} \int_0^\infty c \scs{\rm T}(k,t) {\rm d} t$. We mention that the
precision of calculations of the generalized dielectric permittivity at
low frequencies is very sensitive to statistical uncertainties of data for
total current fluctuations $c \scs{\rm L,T}(k,t)$ obtained in computer
experiment. An additional source of errors is the truncation of long time
tails in correlation functions at numerical integration. As a consequence,
we could not provide a satisfactory reproduction of $\varepsilon \scs{\rm
T}(k,\omega)$ directly in terms of $c \scs{\rm T}(k,t)$, especially at small
wavevectors and in the case of the Ewald method, where the length of the
simulation was twice smaller than for the ISRF geometry. For the last
reason, all other results will be presented in the ISRF geometry only.
The computations show that a much more reliable evaluation can be performed
when the generalized permittivity $\varepsilon \scs{\rm T}(k,\omega)$ is
found via relation (15), i.e., in terms of the transverse dielectric
$\epsilon \scs{\rm T}(k,\omega)$ and magnetic $\mu(k,\omega)$ functions
which correspond to polarization $g \scs{\rm T}(k,t)$ and magnetization
$s \scs{\rm T}(k,t)$ fluctuations. Then in view of (16), relation (15)
transforms at $\omega \to 0$ to
\begin{equation}
\lim_{\omega \to 0} \varepsilon \scs{\rm T}(k,\omega)=
\epsilon \scs{\rm T}(k) -{\rm i} h \frac{c^2 k^2}{\omega}
\lim_{\omega \to 0}{\scr L}_{{\rm i}\omega} (s \scs{\rm T}(k,t))
+ {\cal O}(\omega,c^{-4}) \ ,
\end{equation}
where $\epsilon \scs{\rm T}(k)=\lim_{\omega \to 0} \epsilon \scs{\rm T}
(k,\omega)=1+h g \scs{\rm T}(k)$ is the static dielectric function and
$g \scs{\rm T}(k)=\lim_{t \to 0} g \scs{\rm T}(k,t)$. From equality (18)
we immediately obtain that $\varepsilon' \scs{\rm T}(k)=\epsilon \scs{\rm
T}(k)-h c^2 k^2 \int_0^\infty t s \scs{\rm T}(k,t) {\rm d} t$ and
$\varepsilon'' \scs{\rm T}(k)=4 \pi \sigma \scs{\rm T}(k)/\omega$
with $\sigma \scs{\rm T}(k)=\frac{h}{4\pi} c^2 k^2 \int_0^\infty s
\scs{\rm T}(k,t) {\rm d} t$.

The wavevector-dependent dielectric function $\epsilon \scs{\rm T}(k)$
is shown in fig.~3a by the solid curve. In the infinite-wavelength limit
$k\to0$, this function tends to the value $\varepsilon \scs{0}=\lim_{k\to0}
\varepsilon \scs{\rm L}(k) \approx 53$, corresponding to the usual dielectric
constant in the absence of any spatial and time dispersions. It is equal
to unity in the opposite limit $k\to\infty$, indicating that the system
exhibits no dielectric response with respect to strong spatially
inhomogeneous electric fields. The dielectric function $\epsilon^{\rm PD}
\scs{\rm T}(k)=1+h g^{\rm PD} \scs{\rm T}(k)$, calculated in the point dipole
(PD) approximation $g^{\rm PD} \scs{\rm T}(k)=\langle \hat{\scr P}_{\rm T}
(\bms{k},0) \bms{\cdot} \hat{\scr P}_{\rm T}(-\bms{k},0) \rangle \scs{0}
\Big/ 2Nd^2$, is presented in fig.~3a by the dashed curve. As can be seen
from the figure, this function behaves like that for a Stockmayer fluid
[13]. The PD approach is suitable for the reproduction of $\epsilon \scs{\rm
T}(k)$ at very small wavenumbers only, namely, at $k \ll \pi/r \sim 3.4{\rm
\AA}^{-1}$, where $r=\max_a|\bms{\delta}_i^a| \sim 0.92$\AA\ denotes the
radius of the TIP4P molecule. In this wavevector range $\bms{\hat{\cal P}}
\approx {\hat{\scr P}}$ and the spatial extend of charges within the molecule
can be not taken into account when constructing the operator of microscopic
polarization density. At greater wavevectors, the PD function differs
drastically from that obtained within the exact interaction site description.
In particular, the dielectric function $\epsilon^{\rm PD} \scs{\rm T}(k)$
tends to the wrong Onsager value $1+h/3=17.4$ in the infinite-wavevector
limit $k\to\infty$.

The wavevector-dependent coefficient $\sigma \scs{\rm T}(k)$ of transverse
conductivity has been computed in two ways, namely, using the relations
$\frac{h}{4\pi} \int_0^\infty c \scs{\rm T}(k,t) {\rm d} t$ and
$\frac{h}{4\pi} c^2 k^2 \int_0^\infty s \scs{\rm T}(k,t) {\rm d} t$.
The corresponding results are shown in fig.~3b by open circles and the
solid curve, respectively. These two approaches are mathematically
equivalent, but may lead to different results in numerical calculations.
For instance, the exact infinite-wavelength behaviour $\gamma k^2/4\pi$
of function $\sigma \scs{\rm T}(k)$, where $\gamma=h c^2 \lim_{k\to0}
\int_0^\infty s \scs{\rm T}(k,t) {\rm d} t$, has been reproduced exactly
by us only in the second approach which, therefore, should be considered
as a more preferable method of the calculations. As we can see from the
figure, the coefficient of transverse conductivity $\sigma \scs{\rm T}(k)$
takes nonzero values for arbitrary nonzero wavevectors, contrary to the
longitudinal conductivity.

It is necessary to stress that the transverse dielectric permittivity
$\varepsilon \scs{\rm T}(k,\omega)$ has a singularity when both wavevector
and frequency go to zero. The asymptotic behaviour near the singularity can
be obtained on the basis of equation (18) using the equality $\lim_{k\to0}
\varepsilon' \scs{\rm T}(k)=\lim_{k\to0} \epsilon \scs{\rm T}(k)=\varepsilon
\scs{0}$. Then one finds that $\lim_{k,\omega \to 0} \varepsilon \scs{\rm
T}(k,\omega)=\varepsilon \scs{0} -{\rm i} \gamma k^2/\omega$. From the last
expansion it is easy to see that $\varepsilon \scs{\rm T}(k,\omega)$ is a
discontinuous function and its value depends on the order of the limiting
transitions $k,\omega \to 0$, i.e., $\lim_{\omega \to 0} \lim_{k \to 0}
\varepsilon \scs{\rm T}(k,\omega)=\varepsilon \scs{0}$ but $\lim_{k \to 0}
\lim_{\omega \to 0} \varepsilon \scs{\rm T}(k,\omega)= -{\rm i} \infty$.
As in the case of longitudinal fluctuations, the singularity of $\varepsilon
\scs{\rm T}(k,\omega)$ does not lead to singularities in observing quantities
and does not break the physical requirements imposed on dielectrics. In
particular, the macroscopic current $\bms{I}_{\rm T}(\bms{k},\omega)$ does
not appear at $\omega \to 0$, despite the fact that the coefficient $\sigma
\scs{\rm T}(k,\omega)$ of the polarization conductivity accepts nonzero
values at $k \ne 0$. Indeed, by the definition $\bms{I}_{\rm T}(\bms{k},
\omega)=\sigma \scs{\rm T}(k,\omega) \bms{E}_{\rm T}(\bms{k},\omega)$, where
the transverse electric field can be defined according to the Maxwell
equation via the magnetic field as $\bms{E}_{\rm T}(\bms{k},\omega)=-\frac
{\omega}{ck} \bms{B}(\bms{k},\omega) \bms{\times} \bms{\hat k}$, so that the
current vanishes as $\sim \omega$ at $\omega \to 0$. If $k$ tends to zero
additionally, we obtain, taking into account the asymptotic values $\gamma
k^2 \Big/ 4\pi$ of $\sigma \scs{\rm T}(k,\omega)$, that $\lim_{k,\omega\to0}
\bms{I}_{\rm T}(\bms{k},\omega)=- \gamma k \omega \bms{B}(\bms{k})
\bms{\times} \bms{\hat k} \Big/ 4 \pi c \to 0$, where $\bms{B}(\bms{k})=
\lim_{\omega \to 0} \bms{B}(\bms{k},\omega)$, and the macroscopic current
goes to zero again without any singularities.

At small frequencies, the partial contributions of the transverse component
of dielectric functions into observing quantities are small (proportional
to $\omega$) with respect to the corresponding contributions of the
longitudinal part. That is why the functions $\epsilon \scs{\rm T}(k)$ and
$\sigma \scs{\rm T}(k)$, describing the wavevector dependence of $\epsilon
\scs{\rm T}(k,\omega)$ and $\varepsilon \scs{\rm T}(k,\omega)$ in this
frequency range, have not so important physical meaning as the static
longitudinal dielectric permittivity $\varepsilon \scs{\rm L}(k)$. Moreover,
since static electric fields are purely longitudinal, the response
associated with $\epsilon \scs{\rm T}(k)$ and $\sigma \scs{\rm T}(k)$
cannot be realized phenomenologically in a homogeneous isotropic medium,
except as the limits
$$
\epsilon \scs{\rm T}(k)=\lim_{\omega \to 0} \left[ 1+4\pi\frac{{\cal P}
\scs{\rm T}(k,\omega)} {E \scs{\rm T}(k,\omega)} \right] \ , \ \ \ \ \
\sigma \scs{\rm T}(k)=\lim_{\omega \to 0} \frac{I \scs{\rm T}(k,\omega)}
{E \scs{\rm T}(k,\omega)} \ .
$$
We note also that the function $\varepsilon' \scs{\rm T}(k)$ has no physical
meaning for $k\ne0$ at all. This is so because $\lim_{\omega\to0,k\ne0}
\varepsilon' \scs{\rm T}(k)/\varepsilon'' \scs{\rm T}(k)=0$ and the
wavevector-dependent conductivity $\sigma \scs{\rm T}(k)=\frac{{\rm i}
\omega}{4\pi} \lim_{\omega\to0} (\varepsilon \scs{\rm T}(k,\omega)-1)$
is determined in terms of the imaginary part $\varepsilon'' \scs{\rm T}(k)$
exclusively. From the other hand, in the infinite-wavelength limit $k\to0$
when the conductivity vanishes, $\lim_{k\to0} \sigma \scs{\rm T}(k)=0$,
it is not necessary to consider the real part $\varepsilon' \scs{\rm T}(k)$
as an independent quantity because of $\lim_{k\to0} \varepsilon' \scs{\rm
T}(k)=\lim_{k\to0} \epsilon \scs{\rm T}(k)=\varepsilon \scs{0}$.

The normalized time correlation functions $\Phi_{\rm T}(k,t)=g \scs{\rm
T}(k,t)/g \scs{\rm T}(k)$ and $\Upsilon(k,t)=s \scs{\rm T}(k,t)/s \scs{\rm
T}(k)$, related to dynamical polarization and magnetization fluctuations,
are plotted in figs.~4 and 5, respectively. The librational oscillations
superimposed on the exponential, found previously [1, 6] for longitudinal
polarization fluctuations, are identified for transverse functions $\Phi_{\rm
T}(k,t)$ as well. But the oscillations damp more quickly and their amplitudes
are much smaller in this case. The oscillations vanish completely at larger
wavevectors, namely at $k > 20 k_{\rm min}$. The magnetization correlation
functions $\Upsilon_{\rm T}(k,t)$ also exhibit oscillatory features, which
are more visible at small wavenumbers. These functions, however, can accept
as positive as well as negative values, contrary to the polarization
correlations $\Phi_{\rm T}(k,t)$ which remain positive anywhere in time
space. It is worth remarking that after a sufficiently long period, the
polarization functions decay purely exponentially in time. This fact has
been taken into account by us to calculate the contributions of long tails
into the dielectric function $\epsilon \scs{\rm T}(k,\omega)$ at time
integration (14) of $g \scs{\rm T}(k,t)$. In order to demonstrate again
that the true choice of microscopic variables $\bms{\hat{\cal P}}_{\rm T}$
and $\bms{\hat{\cal M}}_{\rm T}$ for describing polarization and
magnetization fluctuations in ISM fluids is so important, analogous
functions, $\langle \hat{\scr P}_{\rm T}(\bms{k},0) \bms{\cdot} \hat{\scr
P}_{\rm T}(-\bms{k},t) \rangle \scs{0} \Big/ \langle \hat{\scr P}_{\rm T}
(\bms{k},0) \bms{\cdot} \hat{\scr P}_{\rm T}(-\bms{k},0) \rangle \scs{0}$
and $\langle \hat{\scr M}_{\rm T}(\bms{k},0) \bms{\cdot} \hat{\scr M}_{\rm
T}(-\bms{k},t) \rangle \scs{0} \Big/ \langle \hat{\scr M}_{\rm T}(\bms{k},0)
\bms{\cdot} \hat{\scr M}_{\rm T}(-\bms{k},0) \rangle \scs{0}$, obtained in
the PD approximation, are also included in figs.~4 and 5 as dashed curves.
The time correlation functions $\Phi_{\rm T}(k,t)$ and $\Upsilon(k,t)$,
corresponding to the exact interaction site description (equations (12)
and (13)), are identical to PD functions in the infinite-wavelength limit
$k\to0$, but they differ between themselves in a characteristic way at
greater wavevector values, namely, at $k> k_{\rm min}$.

The real $\epsilon'\scs{\rm T}(k,\omega)$ and imaginary $\epsilon'' \scs{\rm
T}(k,\omega)$ parts of the transverse dielectric function $\epsilon \scs{\rm
T}(k,\omega)=\epsilon' \scs{\rm T}(k,\omega) - {\rm i} \epsilon'' \scs{\rm
T}(k,\omega)$ for the TIP4P water as depending on frequency at fixed nonzero
wavevectors are shown in figs.~6 and 7 as solid and dashed curves,
respectively. We mention that $\lim_{k \to 0} \epsilon \scs{\rm T}(k,\omega)=
\lim_{k \to 0} \varepsilon \scs{\rm L,T}(k,\omega)=\varepsilon(\omega)$. For
the purpose of comparison the corresponding result obtained previously [1]
for the longitudinal dielectric permittivity $\varepsilon \scs{\rm L}(k,
\omega)$ is also included in fig.~6. The frequency dependence of the
transverse dielectric function $\epsilon \scs{\rm T}(k,\omega)$ at low
enough $\omega$ can be described by the Debye theory for arbitrary
wavevectors. This is a result of the fact that the polarization correlation
functions $g \scs{\rm T}(k,t)$ damp exponentially at long times. With
increasing frequency the collective molecular librations take a prominent
role in forming the frequency shape for $\epsilon \scs{\rm T}(k,\omega)$,
especially at small and intermediate wavevector values when $k \le 20
k_{\rm min}$. In this wavenumber region, above some $\omega(k)$=1-10 THz
corresponding to the position of the first maximum of $\epsilon''\scs{\rm
T}(k,\omega)$, the relaxation behaviour changes into a librational resonance
process, characterizing by a frequency of order 100 THz. This frequency is
associated with the position of the second maximum of $\epsilon''\scs{\rm
T}(k,\omega)$, which practically does not depend on wavevector. At
sufficiently great frequencies the transverse component tends to the
longitudinal dielectric constant, so that, for example, at small wavenumbers
$k \le 2 k_{\rm min}$ the both components become indistinguishable from one
another at $\omega > 10$ THz (see fig.~6). For larger wavevectors the
transverse function (fig.~7) differs from the longitudinal one (figs.~6, 7
of [1]) considerably. Beginning from frequencies of order $\omega \sim 1000$
THz and wavevectors of order $k \sim 100 {\rm \AA}^{-1}$ the transverse
dielectric function tends to the limiting value $\varepsilon_\infty =1$
corresponding to nonpolarizable systems.

\subsection{Magnetic properties}

The wavevector- and frequency-dependent magnetic susceptibility \mbox
{$\chi(k,\omega)\!=\!\mu(k,\omega)\!-\!1$} $=-\chi'(k,\omega) - {\rm i}
\chi''(k,\omega)$ has been evaluated for the TIP4P water using fluctuation
formula (16). Its real $\chi'(k,\omega)$ and imaginary $\chi''(k,\omega)$
parts are plotted in fig.~8 as functions of frequency at fixed wavevectors
by solid and dashed curves, respectively. We note that fluctuation formula
(16) is somewhat other by the structure than fluctuation formulas (2) and
(14) for dielectric functions $\varepsilon \scs{\rm L}(k,\omega)$ and
$\epsilon \scs{\rm T}(k,\omega)$. That is why, unlike the dielectric
susceptibility $\epsilon \scs{\rm L,T}(k,\omega)-1$, the magnetic
susceptibility $\chi(k,\omega)$ vanishes in the static limit at arbitrary
wavenumbers, $\lim_{\omega \to 0} \chi(k,\omega) \equiv \chi(k)=0$, and
this statement follows directly from equation (16). From the physical point
of view, such a situation is explained by the different nature of electric
and magnetic dipoles in ISM fluids. While the electric dipole moment
$\bms{d}_i$ is fixed by the rigid molecular geometry, resulting in
$|\bms{d}_i|=d=$const, the magnetic moment $\bms{m}_i$ is caused by the
rotational motion of charges sites and it depends explicitly on the angular
velocity $\bms{\mit \Omega}_i$ of the molecule. Thus, the vector $\bms{m}_i$
varies in time not only due to the orientational dynamics of the molecule,
as $\bms{d}_i$-vector, but also owing to the changes of $\bms{\mit
\Omega}_i$, so that $|\bms{m}_i|\ne$const. As a result, the macroscopic
magnetization does not appear in the presence of static magnetic fields
$\bms{H}_0$, because then $\langle \bms{\mit \Omega}_i \rangle=0$ even
if the spatial inhomogeneity of $\bms{H}_0$ is taken into account
(quasiequilibrium, stationary state of the system).

At nonzero frequencies, when the system is far from equilibrium ($\langle
\bms{\mit \Omega}_i \rangle \ne 0$) due to the presence of external timely
inhomogeneous fields, two mechanisms of appearing the macroscopic
magnetization are possible. The first mechanism, is connected with the
alignment of own magnetic dipole moments along the field $\bms{H}_0$,
leading to a paramagnetic-like behaviour with $\chi'(k,\omega)>0$. The
second one is caused by the magnetizability of molecules owing to the
changes of angular velocities in magnetic fields. Then the corresponding
additional magnetic moment of the system will be directed oppositely to
$\bms{H}_0$-vector that may lead to a diamagnetic-like behaviour with
$\chi'(k,\omega)<0$. As we can see from fig.~8, at relatively small
frequencies, the TIP4P water exhibits paramagnetic features. Beginning
from frequencies of order 100-300 THz (in dependence on wavevector) and
higher the diamagnetic contributions into the magnetization become to
dominate and the system under consideration behaves like a diamagnetics.
It is interesting to remark that the frequencies $\omega_{\rm p}$ and
$\omega_{\rm d}$, corresponding to maximal values of $|\chi'(k,\omega)|$
in the paramagnetic and diamagnetic regions, respectively, almost do not
depend on wavevector in a wide wavenumber range from $k=0$ up to $k \le
20 k_{\rm min}$, where they are equal to $\omega_{\rm p} \sim 80$ THz and
$\omega_{\rm d} \sim 180$ THz, indicating about a weak influence of the
translational motions on the processes of magnetization.

Applying the Laplace boundary theorem to fluctuation formula (16) yields
the magnetic susceptibility $\chi_\infty(k)=\lim_{\omega \to \infty}
\chi(k,\omega)=-h s \scs{\rm T}(k) \le 0$ in the infinite-frequency regime,
where $s \scs{\rm T}(k)= \lim_{t\to0} s \scs{\rm T}(k,t)>0$ is the static
autocorrelation function. The function $\chi_\infty(k)$ is plotted in
fig.~9 by the solid curve. It takes negative values at arbitrary finite
wavenumbers and tends to zero as far as $k\to\infty$ (note that the
magnetic susceptibility is shown in figs.~8 and 9 for convenience in the
negative representation $-\chi$). The existence of a nonzero coefficient for
the magnetic susceptibility in the infinite-frequency regime may be
considered as a somewhat unexpected result. For example, the dielectric
susceptibilities $\epsilon \scs{\rm L,T}(k,\omega)-1$ vanish (as $\sim 1/
\omega^2$) at $\omega\to \infty$, because electric dipoles are unsensitive
to very fast changes of finite electric fields in time owing to the inertness
of molecules. For the case of magnetic susceptibility the pattern is
different for the following reason. According to material relations, the
macroscopic magnetization of the system can be presented in the form
$\bms{\cal M}_{\rm T}(\bms{k},\omega)=\frac{1}{4\pi} \frac{\chi(k,\omega)}
{1+\chi(k,\omega)} \bms{B}_{\rm T}(\bms{k},\omega)$. The vector of magnetic
induction is not independent at $k,\omega \ne 0$ and expressed in terms of
the electric field using the Maxwell equation as $\bms{B}_{\rm T}(\bms{k},
\omega) = - \frac{c}{\omega} \bms{k} \bms{\times} \bms{E}({\bms{k},\omega})$.
So that in the infinite-frequency regime the macroscopic magnetization
$\bms{\cal M}_{\rm T}(\bms{k},\omega)$ vanish as $1/\omega$ at finite values
of electric fields $\bms{E}({\bms{k},\omega})$, despite the fact that
$\chi_\infty(k) \ne 0$. A nonvanishing magnetization of the system can be
achieved at $\omega\to\infty$ for infinite values ($\bms{E}({\bms{k},\omega})
\sim \omega$) of electric fields only. But this corresponds rather a
hypothetical case which is hard to realize in practice. Moreover, it
cannot be handled within the linear response theory, where the
smallness of electromagnetic fields is assumed in advance.

The value of function $\chi_\infty(k)$ in the infinite-wavelength limit
can be presented in an analytical form. Acting in the spirit of derivation
of analytical formulas for the static current correlations (see Appendix)
and taking into account explicit expression (13) for the microscopic
magnetization density $\bms{\hat{\cal M}}_{\rm T}$, one obtains the
following result
\begin{equation}
\chi_\infty=\lim_{k\to0} \chi_\infty(k)=-\frac13 \frac{\pi N}{c^2 V}
\sum_{a,b}^M q \scs{a} q \scs{b} \sum_{\alpha,\beta}^{X,Y,Z}
\left( \left[ \frac{1}{J}-\frac{2}{J_\alpha} \right]
(\Delta_a^\alpha \Delta_b^\beta)^2+\frac{1}{J_\alpha}
\Delta_a^\alpha \Delta_a^\beta \Delta_b^\alpha \Delta_b^\beta \right) \ ,
\end{equation}
where $J_\alpha$ are the moments of inertia of the molecule with respect to
its three principal axes $X,Y,Z$, the $\alpha$th component of vector-position
$\bms{\delta}_i^a$ for site $a$ in the molecular principal coordinate system
is denoted as $\Delta_a^\alpha$ and $1/J=1/J_X+1/J_Y+1/J_Z$. It can be shown
easily that in the PD approximation the function $\chi_\infty^{\rm PD}(k)=
-h \langle \hat{\scr M}_{\rm T}(\bms{k},0) \bms{\cdot} \hat{\scr M}_{\rm
T}(-\bms{k},0) \rangle \scs{0} \Big/ 2Nd^2$ does not depend on wavevector,
i.e., $\chi_\infty^{\rm PD}(k)=\chi_\infty$ (the horizontal dashed line in
fig.~9). The values of $\chi_\infty^{\rm PD}(k)$ computed in the MD
calculations are shown in the figure as open circles. They differ from the
exact value $\chi_\infty$ within statistical noise. Using equation (19) we
obtain for the TIP4P model: $\chi_\infty \approx -6.7 \cdot 10^{-10}$ that
is too small in comparison with the diamagnetic susceptibility $\chi \scs
{\rm H_20} \approx -9 \cdot 10^{-6}$ of real water. This is so because
ISMs do not take into account an electronic subsystem which gives the
main contribution into the magnetic susceptibility owing to the smallness
of the mass $m_{\rm e}$ of electron. This contribution can be estimated
noticing that the diamagnetic susceptibility of an ideal electron gas is
defined as $\chi_{\rm e} \sim - N e^2 \langle \rho_{\rm e}^2 \rangle\Big/V
m_{\rm e} c^2$, where $e$ is the electron charge and $\langle \rho_{\rm e}^2
\rangle$ denotes the averaged value of square radius-vectors of electrons
within the molecule [24]. Comparing this result with formula (19) and
taking into account that $\langle \rho_{\rm e}^2 \rangle \sim r^2$, $|q
\scs{a}| \sim |e|$, $\max_\alpha[1/J_\alpha] \sim 1/m \scs{\rm H} r^2$ and
${\bms{\Delta}_a}^2 \sim r^2$ yields $\chi_\infty/\chi_{\rm e} \sim m_{\rm
e}/m \scs{\rm H} \sim 10^{-4} \sim \chi_\infty/\chi \scs{\rm H_20}$, where
$m \scs{\rm H}$ is the mass of a hydrogen atom. It is necessary to point out
that the atomic diamagnetic susceptibility (19) in the infinite-frequency
limit, like the case of electronic diamagnetism, does not depend on
temperature of the system and it is determined by the geometry of the
molecule. We note that such an atomic diamagnetism can be absent for
some particular molecular geometries. For example, for the simplest $\xi$DS
model, when $M=2$, $q \scs{1}=q$, $q \scs{2}=-q$, $\bms{\Delta}_1=\bms{l}$
and $\bms{\Delta}_2=-\bms{l}$, we find from equation (19) that
$\chi_\infty^{\xi \rm DS}=0$.

We would like also to emphasize that fluctuation formula (16), which
explicitly involves the microscopic magnetization density, is only the way
to calculate the magnetic susceptibility $\chi(k,\omega)$ of ISM fluids
in computer experiment not only for nonzero wavevectors but also at
$k=0$. Such a numerical calculation cannot be done within the abbreviated
description, despite the fact that according to relation (17), the knowledge
of the longitudinal $\varepsilon \scs{\rm L}(k,\omega)$ and transverse
$\varepsilon \scs{\rm T}(k,\omega)$ components of the generalized dielectric
permittivity (equations (2) and (6)) allows one, in principle, to determine
the frequency-dependent magnetic susceptibility $\chi(\omega)$ in the
infinite-wavelength limit $k\to0$. The reasons for this situation are
following. First of all we underline that relation (17) is valid for nonzero
but very small values of wavevector, namely, for $k \mathop{_\sim} \limits^{
{\mbox{\footnotesize $<$}}} \omega /c$. Even for relatively great frequencies
of order 1000 THz this condition corresponds to $k \mathop{_\sim} \limits^
{{\mbox{\footnotesize $<$}}} 0.0003{\rm \AA}^{-1}$. At the same time, the
smallest nonzero value of wavevectors accessible in our simulations is
$k_{\rm min} \sim 0.3{\rm \AA}^{-1}$. So that to reach the value $0.0003{\rm
\AA}^{-1}$ it is necessary to increase the size of the simulation box to
1000 times! From the other hand, at $k = k_{\rm min}$ we can use relation
(17) beginning from a frequency of order $\omega \sim c k_{\rm min} \sim
10^6$ THz, where $\varepsilon \scs{\rm T}(k_{\rm min},\omega) \approx
\chi_\infty(k_{\rm min})$. But it is known in advance that the magnetic
susceptibility of the system is too small, consisting, for instance,
approximately $-7.5 \cdot 10^{-10}$ for $\chi_\infty(k)$ at $k=k_{\rm min}$.
Therefore, to reproduce this value using the difference of functions
$\varepsilon \scs{\rm L}(k_{\rm min},\omega)$ and $\varepsilon \scs{\rm T}
(k_{\rm min},\omega)$ it is required to compute them with a relative
statistical accuracy of order $10^{-12}$ at least that constitutes an
unrealistic problem again. For example, even at extra long simulations
with 1 000 000 time steps, as in our case, the relative statistical
uncertainties for the dielectric quantities are of order 1\%, i.e.,
$10^{-2}$ only.

\subsection{Propagation of electromagnetic waves}

We consider now the question of propagation of free transverse
electromagnetic waves, $E(\bms{r},t) \sim {\mbox{\large e}}^{{\rm i}(\omega
t-\bvs{k\!\cdot\!r})}$, in the TIP4P water. The dispersion relation,
connecting frequency and wavenumber of the waves in dielectrics, is of
the form [18]:
\begin{equation}
\omega^2 - \frac{c^2 k^2}{\varepsilon \scs{\rm T}(k,\omega)} = 0 \ .
\end{equation}
In this case $\omega \sim ck$ and, as it follows from equation (15), the
transverse functions $\varepsilon \scs{\rm T}(k,\omega)$ and $\epsilon
\scs{\rm T}(k,\omega)$ differ from one another by order of $\chi(k,\omega)$.
As was shown in the preceding subsection, the magnetic susceptibility of
ISM systems is sufficiently small at arbitrary wavevectors and frequencies,
i.e., $\chi(k,\omega) \ll 1$. Thus we can putt $\varepsilon \scs{\rm T}(k,
\omega) \approx \epsilon \scs{\rm T}(k,\omega)$ in relation (20) without loss
of precision. Further, the characteristic frequency scale of varying the
dielectric constant is of order 1000 THz. This corresponds to an interval
of varying wavenumbers in the waves of order $0.0003{\rm \AA}^{-1}$. At the
same time, the characteristic wavevector scale of varying $\epsilon \scs{\rm
T}(k,\omega)$ is of order $k_{\rm min} \sim 0.3{\rm \AA}^{-1}$, so that the
spatial dispersion of the dielectric constant in equation waves (20) can be
neglected completely, i.e., $\varepsilon \scs{\rm T}(k,\omega) \approx
\epsilon \scs{\rm T}(k,\omega) \approx \varepsilon(\omega)$. The dielectric
permittivity $\varepsilon(\omega)$ is an imaginary function of frequency.
Therefore, a solution to the equation (20) with respect to wavenumbers at
a given real frequency (it corresponds to frequency of the source which
creates electromagnetic waves) it is necessary to find in the form $\bms{k}=
\bms{k'}-{\rm i} \bms{k''}$, where the magnitude $k''$ of the imaginary part
will describe the damping of waves with the wavelength $\lambda=2\pi/k'$. We
shall consider a case when plane waves are damped in direction of their
propagation, i.e., when $\bms{k'} \bms{\cdot} \bms{k''}=k' k''$.

Numerical results for the dispersion $\omega(k')$ and damping $k''(\omega)$,
coefficients in the infrared region of spectrum are shown in figs.~10a and
10b, respectively. We note that calculation of the damping coefficient at
great frequencies is very sensitive to the precision of evaluation of the
dielectric constant. A more accurate estimation can be achieved when the
dielectric constant is determined in terms of current fluctuations via
fluctuation formula (5) (the solid curve in fig.~10b), instead of the usual
formula (2) (the open circles). As far as $k'' \ne 0$ in the whole region
of frequencies $0 < \omega < \infty$, purely monochromatic waves cannot
propagate in dielectrics. Nevertheless, choosing a criterion $k'' \ll k'$,
we can talk about quasimonochromatic waves with a slight absorption.
According to our calculation, only two opposite regions of very small
($\omega \mathop{_\sim} \limits^{{\mbox {\footnotesize $<$}}} 0.01$ THz;
radiowaves) and very great ($\omega \mathop{_\sim} \limits^{{\mbox
{\footnotesize $>$}}} 500$ THz; far infrared, visible light) frequencies
can satisfy this criterion. In the radiowaves region (see insets of the
figures), the phase velocity $\omega(k')/k'=c/\sqrt{\varepsilon_0}$ is
defined by the static dielectric constant $\sqrt{\varepsilon_0} \approx 7$
(for real water $\sqrt{\varepsilon_0} \approx 9$). For example, at $\omega
\approx 0.01$ THz we obtain $k'c \approx 0.1$ THz and $k''c \approx 0.002$
THz. Thus, the radiowaves with the wavelength $\lambda=2\pi/k' \approx 2$cm
are dumped in a characteristic way during the interval $2\pi/k'' \approx 1$m.
In the far infrared region the phase velocity of the nonpolarizable TIP4P
water tends to its infinite frequency value $c/\sqrt{\varepsilon_\infty}$,
where $\varepsilon_\infty=\lim \limits_{\omega \to \infty} \varepsilon
(\omega)=1$. Owing to the electronic polarizability of molecules, which is
not taken into account in the TIP4P model, the dielectric constant of real
water differs from unity even in the visible light spectrum $\omega \sim
3 000$ THz, where $\sqrt{\varepsilon(\omega)} \approx 1.33$. It is
interesting to remark about the existence of a large region of frequencies
($\omega \sim 10-50$ THz), where the group velocity $\partial \omega(k')/
\partial k'=c/\sqrt{\varepsilon^*}$ is practically constant. In this region,
near the librational minimum of the imaginary part $\varepsilon''(\omega)$,
the real part $\varepsilon'(\omega)$ of the dielectric constant approximately
does not depend on frequency and takes values of order $\varepsilon^* \approx
2.3$ (see fig.~6a of Ref.~[1] for $\varepsilon(\omega)$ and fig.~6a of
the present paper for $\epsilon \scs{\rm T}(k_{\rm min},\omega) \approx
\varepsilon(\omega)$). However, the waves are damped significantly ($k''
< k'$) in this frequency range. Within the absorption maximum, which
occurs at $\sim 150$ THz, we have $k'' \sim k'$, so that electromagnetic
waves decay here on an interval of order of their wavelengths.

All our previous calculations of the dielectric functions $\varepsilon
\scs{\rm L,T}(k,\omega)$ dealt with real values of wavevector and frequency.
Such functions at a given single set of $k$ and $\omega$ describe the
response of the system on electromagnetic fields in the form of monochromatic
plane waves $\,\sim {\mbox{\large e}}^{{\rm i}(\omega t-\bvs{k\!\cdot\!r})}$.
Imaginary values of frequency and wavevector correspond to cases when
amplitudes of fields either increase or decrease in time and space. Since
arbitrary inhomogeneous fields can be cast as a set of the monochromatic
waves, using the time and spatial Fourier transform, the dielectric function
at imaginary values of $k$ and $\omega$ may be expressed in terms of its
values at real wavevectors and frequencies. As a demonstration, we consider
the case of purely imaginary frequencies, $\omega \equiv {\rm i} \omega^*$,
in the infinite wavelength limit ($k=0$). Then, as it follows from
fluctuation formula (2), the dielectric constant can be calculated directly
\begin{equation}
\frac{\varepsilon(\omega^*)-1}{\varepsilon \scs{0}-1}=\left(
1+\omega^* \int \limits_0^\infty {\mbox{\large e}}^{\omega^*t}
\Phi(t) {\rm d} t \right) \ ,
\end{equation}
using the time correlation function $\Phi(t)=\left< \sum_{i,j}^N \bms{d}_i(0)
\bms{\cdot} \bms{d}_j(t) \right> \scs{0} \Big/ \left< \sum_{i,j}^N \bms{d}_i
(0) \bms{\cdot} \bms{d}_j(0) \right> \scs{0}$ of the total dipole moment.
From the other hand, presenting the field $\,\sim {\mbox{\large e}}^
{-\omega^*t}$ in the form of the corresponding Fourier integral over real
frequencies, it can be shown, in particular, that at $\omega^*< 0$ the
dielectric constant $\varepsilon(\omega^*)$ is expressed via the imaginary
part of $\varepsilon(\omega)$ (defined at real frequencies) as follows [17]:
\begin{equation}
\varepsilon(\omega^*)-1=\frac2\pi \int \limits_0^\infty
\frac{x \varepsilon''(x)}{x^2+{\omega^*}^2} {\rm d} x \ .
\end{equation}
The function $\varepsilon(\omega^*)$ is plotted in fig.~11a. It describes the
response of the system on external electric fields which increase in time
exponentially. As was expected, this function at $\omega^*\to -\infty$ tends
to unity owing the inertness of polar molecules.

The case $\omega^*> 0$ will correspond to decaying in time of electric
fields. This decaying can be caused either by decreasing of external fields
or by switching off of sources which support these fields. As an example,
consider the spatial and time decay of electromagnetic fields in the system
after the passage of transverse electromagnetic waves. In this situation,
frequency and wavevector are purely imaginary quantities, i.e., $\omega
\equiv {\rm i}\omega^*$ with $\omega^*> 0$ and $k \equiv -{\rm i} k''$, that
is necessary to take into account finding a solution to the equation (20).
For the same reasons as in the case of propagation of electromagnetic waves,
we put $\varepsilon \scs{\rm T}(k,\omega) \approx \varepsilon(\omega)$ in
equation (20) and obtain $\omega^*=ck''/\sqrt{\varepsilon(\omega^*)}$, where
$\varepsilon(\omega^*)$ accepts purely real values at purely imaginary
frequencies. We note that an independent parameter in this equality is $k''$,
so that the function $\omega^*(k'')$ will describe the time decay $\,\sim
{\mbox{\large e}}^{-\omega^*t}$ of the transverse electric field with
a spatial inhomogeneity of $k''$ which has been created by the passed
electromagnetic wave (see fig.~10b). The function $\omega^*(k'')$ is shown
in fig.~11b. As can be seen, weak spatial inhomogeneities correspond to long
life times $\,\sim 1/\omega^*(k'')$, whereas strong inhomogeneous fields damp
in time faster. It is worth to remark also that in the asymptotic limit $t
\to \infty$, the time correlation function $\Phi(t)$ decays in time
exponentially as $\sim {\mbox{\large e}}^{-t/\tau_{\rm rel}}$, where
$\tau_{\rm rel}=6.7$ps is the relaxation time. Therefore, the dielectric
constant $\varepsilon(\omega)$ at imaginary frequencies can be defined only
in the region $\Im \omega < 1/\tau_{\rm rel}=0.149$ THz, because otherwise
the integral in (21) is divergent. This merely means that transverse
electromagnetic excitations cannot damp in time faster than with the
characteristic interval $\tau_{\rm rel}$. The limiting region of imaginary
frequencies is shown in fig.~11b by the horizontal dashed line.

Finally, as far as real $\varepsilon'\scs{\rm T}$ and imaginary
$\varepsilon''\scs{\rm T}$ parts of the transverse wavevector- and
frequency-dependent dielectric constant are defined according to the
fluctuation formula (6) via the same time correlation function $c \scs
{\rm T}(k,t)$, they are not independent and must be connected between
themselves. The desired relations can be obtained using analytical
properties of the functions $c \scs{\rm T}(k,t)$, $\varepsilon'\scs{\rm
T}(k,\omega)$ and $\varepsilon''\scs{\rm T}(k,\omega)-4\pi \sigma \scs
{\rm T}(k)/\omega$. The result is
\begin{equation}
\varepsilon'\scs{\rm T}(k,\omega)-1=\frac2\pi \int \limits_0^\infty
\hspace{-9.32pt}\mbox{\large \bf - }
\frac{x \varepsilon''\scs{\rm T}(k,x)}{x^2-\omega^2} {\rm d} x \ , \ \ \ \ \
\varepsilon''\scs{\rm T}(k,\omega)=-\frac{2\omega}\pi \int \limits_0^\infty
\hspace{-9.32pt}\mbox{\large \bf - }
\frac{\varepsilon'\scs{\rm T}(k,x)-1}{x^2-\omega^2} {\rm d} x
+\frac{4\pi \sigma \scs{\rm T}(k)}{\omega} \ .
\end{equation}
The relations (23) are similar to the well-known Kramers-Kronig expressions
for the dielectric constant of conductors [17]. They can be applied to the
transverse frequency-dependent dielectric constant of ISMs at arbitrary
values of wavevector.

\section{Conclusion}

We have established that the calculation of the dielectric quantities for
interaction site models of polar fluids can be performed in computer
experiment by two alternative ways, namely, using either charge or current
fluctuations. The first way is more efficient to evaluate the longitudinal
component $\varepsilon \scs{\rm L}(k,\omega)$ of the dielectric permittivity,
whereas the second method is suitable to obtain its transverse part
$\varepsilon \scs{\rm T}(k,\omega)$. Separating the total current of moving
charges into electrical and magnetic components, the transverse generalized
dielectric permittivity $\varepsilon \scs{\rm T}(k,\omega) = \epsilon
\scs{\rm T}(k,\omega) + \frac{c^2 k^2}{\omega^2} \frac{\chi(k,\omega)}
{1+\chi(k,\omega)}$ has been expressed in terms of the transverse dielectric
$\epsilon \scs{\rm T}(k,\omega)$ and magnetic $\chi(k,\omega)$ functions.
These functions have been evaluated for the TIP4P model of water by
molecular dynamics simulations in the whole region of wavevectors and
frequencies for the first time on the basis of the proposed fluctuation
formulas.

It has been shown that the transverse $\epsilon \scs{\rm T}(k,\omega)$
and longitudinal $\epsilon \scs{\rm L}(k,\omega) \equiv \varepsilon
\scs{\rm L}(k,\omega)$ components of the dielectric tensor $\bms{\epsilon}
(\bms{k},\omega)$ differ between themselves in a characteristic way and
coincide with one another in opposite limits of either very small wavenumbers
or very large wavevector and frequency values. Moreover, contrary to the
case of longitudinal fluctuations, the generalized dielectric permittivity
$\varepsilon \scs{\rm T}(k,\omega)$ exhibits a specific behaviour when both
wavenumber and frequency tend to zero. We has identified also that at great
frequencies the TIP4P water can be considered as a weak diamagnetics.
However, values of the magnetic susceptibility $\chi(k,\omega)$ are much
smaller in amplitude than those for real water, because the electronic
magnetization is not included in this model. It is worth to stress that in
the present investigation the polarization and magnetization microscopic
densities have been constructed taking into account explicitly the atomic
structure of ISMs. As a result, it has been demonstrated, in particular,
that at nonzero wavevectors the genuine transverse dielectric function
$\epsilon \scs{\rm T}(k,\omega)$ has nothing to do with that obtained in
the point dipole approximation. The last function behaves like the
dielectric permittivity of a Stockmayer system [12--14] and can be
used at small wavenumbers only as an estimation of the frequency-dependent
dielectric constant in the infinite-wavelength limit.

The longitudinal and transverse components of the wavevector- and
frequency-depen\-dent dielectric permittivity describe all electromagnetic
phenomena in the system. The knowledge of these quantities may present an
interest in both theory and pure experiment. The performed calculations can
be extended, in principle, to more realistic models of polar fluids. This
will be the subject of a separate consideration.

\vspace{24pt}

The author would like to acknowledge financial support by the President
of Ukraine.

\newpage
{\small
\begin{center}
{\large \bf Appendix}
\end{center}
\setcounter{equation}{0}
\renewcommand{\theequation}{A\arabic{equation}}

We shall show that the longitudinal and transverse components $c \scs{\rm
L,T}(k)$ of the static wavevec\-tor-depended Kirkwood factor of second order
$c(k)=c \scs{\rm L}(k) + 2 c \scs{\rm T}(k)$ are presented analytically.
According to definitions (3) and (7), the functions $c(k)$ and
$c \scs{\rm L}(k)$ read
\begin{eqnarray}
&&c(k) = \frac{1}{N d^2}  \left< \sum_{i, a}^{N,M} q \scs{a}
\bms{V}_i^a \, \bms{\cdot} \sum_{j, b}^{N,M} q \scs{b}
\bms{V}_j^b \ {\mbox{\large e}}^{-{\rm i} \bvs{k}
\bvs{\cdot} (\bvs{r}_i^a-\bvs{r}_j^b)} \right>_0 \ , \nonumber \\ [-4pt]
\\ [-4pt]
&&c \scs{\rm L}(k) = \frac{1}{N d^2}  \left< \sum_{i, a}^{N,M} q \scs{a}
(\bms{\hat k} \bms{\cdot} \bms{V}_i^a ) \sum_{j, b}^{N,M} q \scs{b}
(\bms{\hat k} \bms{\cdot} \bms{V}_j^b ) \ {\mbox{\large e}}^{-{\rm i} \bvs{k}
\bvs{\cdot} (\bvs{r}_i^a-\bvs{r}_j^b)} \right>_0 \ , \nonumber
\ \ \ \ \ \ \ \ \ \
\end{eqnarray}
where the site velocities can be split as $\bms{V}_i^a=\bms{V}_i+\bms{\mit
\Omega}_i \bms{\times} \bms{\delta}_i^a$. We mention that $\bms{V}_i$ and
$\bms{\mit \Omega}_i$ are the translational and angular velocities of the
$i$th molecule, respectively, $\bms{\delta}_i^a=\bms{r}_i^a-\bms{r}_i$ and
$\bms{r}_i$ denotes the position of the molecular centre of mass.

Equilibrium distribution functions are factored into the coordinate and
velocity parts. In its turn, translational and angular velocities are
distributed independently of one another for each molecule. As a result,
nonzero contributions to equilibrium averaging give only terms with
coincident molecular indexes $(i=j)$ of summation (A1). It is more
convenient to consider each molecule in its own principal coordinate
system in which $\bms{\delta}_i^a \equiv \bms{\Delta}_a$. Then
expressions (A1) transform into
\begin{eqnarray}
&&c(k) = \frac{1}{N d^2} \sum_i^N \sum_{a, b}^{M} q \scs{a}
q \scs{b} \left< \left( \bms{V}_i^2 +
[\bms{\mit \Omega_i} \bms{\times} \bms{\Delta}_a] \bms{\cdot}
[\bms{\mit \Omega_i} \bms{\times} \bms{\Delta}_b] \right)
{\mbox{\large e}}^{-{\rm i} \bvs{k} \bvs{\cdot} \bvs{\rho}_{ab}}
\right>_0 \ , \nonumber \\ [-4pt] \\ [-4pt]
&&c \scs{\rm L}(k) = \frac{1}{N d^2} \sum_i^N \sum_{a, b}^{M} q \scs{a}
q \scs{b} \left< \left( (\bms{\hat k} \bms{\cdot} \bms{V}_i)^2 +
(\bms{\hat k} \bms{\cdot}[\bms{\mit \Omega_i} \bms{\times} \bms{\Delta}_a])
(\bms{\hat k} \bms{\cdot}[\bms{\mit \Omega_i} \bms{\times} \bms{\Delta}_b])
\right) {\mbox{\large e}}^{-{\rm i} \bvs{k} \bvs{\cdot} \bvs{\rho}_{ab}}
\right>_0 \ , \nonumber \ \ \ \ \ \ \ \ \ \
\end{eqnarray}
where $\bms{\rho}_{ab}=\bms{\Delta}_a-\bms{\Delta}_b$. It is necessary to
underline that for a given molecular geometry, $\bms{\Delta}_a$ ($a=1,
\ldots,M$) constitute the set of constant vectors characterizing the
positions of charged sites in the principal coordinate system attached to
the molecule. Further, we use the following equalities $3 (\bms{\hat k}
\bms{\cdot}[\bms{\mit \Omega} \bms{\times} \bms{\Delta}_a]) (\bms{\hat k}
\bms{\cdot}[\bms{\mit \Omega} \bms{\times} \bms{\Delta}_b]) = (3 \bms{\hat k}
\bms{\hat k} - \bms{1})\, \bms{:}\, \bms{\mit \Omega} \bms{\times}
\bms{\Delta}_a \bms{\mit \Omega} \bms{\times} \bms{\Delta}_b + [\bms{\mit
\Omega} \bms{\times} \bms{\Delta}_a] \bms{\cdot} [\bms{\mit \Omega}
\bms{\times} \bms{\Delta}_b]$ and $[\bms{\mit \Omega} \bms{\times}
\bms{\Delta}_a] \bms{\cdot} [\bms{\mit \Omega} \bms{\times} \bms{\Delta}_b]
={\mit \Omega}^2 (\bms{\Delta}_a \bms{\cdot} \bms{\Delta}_b)-(\bms{\mit
\Omega} \bms{\cdot} \bms{\Delta}_a) (\bms{\mit \Omega} \bms{\cdot}
\bms{\Delta}_b)$ and take into account the relations
\begin{equation}
\left< {\mbox{\large e}}^{-{\rm i} \bvs{k} \bvs{\cdot} \bvs{\rho}}
\right>_{\bvs{\hat k}} = j_0(k \rho) \ , \ \ \ \ \ \ \
\left< (3 \bms{\hat k} \bms{\hat k} - \bms{1})
{\mbox{\large e}}^{-{\rm i} \bvs{k} \bvs{\cdot} \bvs{\rho}}
\right>_{\bvs{\hat k}} =
-j_2(k \rho) (3 \bms{\hat \rho} \bms{\hat \rho} - \bms{1}) \ ,
\end{equation}
where $j \scs{0}(z)=\sin(z)/z$, $j \scs{2}(z)=3j \scs{1}(z)/z-j \scs{0}(z)$
are the spherical Bessel functions of order zero and two, respectively,
$\bms{\hat \rho}=\bms{\rho}/\rho$ and averaging in (A3) is performed over
orientations of $\bms{\hat k}$-vector with respect to the molecule. Owing
identity of molecules and isotropy of the system we have, in particular,
$\langle V_i^2 \rangle=\langle V^2 \rangle$, $\langle {\mit \Omega}_i^2
\rangle=\langle {\mit \Omega}^2 \rangle$ and $\langle (\bms{\hat k} \rangle
\bms{\cdot} \bms{V})^2 \rangle=\langle V^2 \rangle /3$. Using the definition
of temperature and the equipartition theorem yields $m \langle V^2 \rangle_0
/3 = J_\alpha \langle {\mit \Omega}_\alpha^2 \rangle_0=k_{\rm B} T$, where
$m$ is the mass of the molecule and ${\mit \Omega}_\alpha$ are the principal
components ($\alpha=X,Y,Z$) of angular velocity. Finally, in view of the
statistical independence of angular velocities directed along different
principal axes of inertia, after cumbersome but not complicate operations
one obtains
\vspace{4pt}
\begin{eqnarray}
c(k) &=& \frac{k_{\rm B}T}{d^2} \sum_{a, b}^{M} q \scs{a}
q \scs{b} j \scs{0} (k \rho_{ab}) \Bigg[ \frac{3}{m}+
\sum_\alpha \left( \frac{1}{J}-\frac{1}{J_\alpha} \right)
\Delta_a^\alpha \Delta_b^\alpha \Bigg] \, , \\ [9pt]
c \scs{\rm L}(k) &=& \frac13 c(k) - \frac{k_{\rm B}T}{3 d^2}
\sum_{a \ne b}^{M} q \scs{a} q \scs{b} j \scs{2} (k \rho_{ab})
\sum_{\alpha, \beta} h_{ab}^{\alpha \beta} \left( 3 \hat \rho_{ab}^\alpha
\hat \rho_{ab}^\beta - \delta_{\alpha \beta} \right) , \nonumber
\\ [-6pt] \\ [-6pt]
c \scs{\rm T}(k) = \frac{c(k)-c \scs{\rm L}(k)}{2}&=&\frac13 c(k) +
\frac{k_{\rm B}T}{6 d^2} \sum_{a \ne b}^{M} q \scs{a} q \scs{b}
j \scs{2} (k \rho_{ab}) \sum_{\alpha, \beta} h_{ab}^{\alpha \beta}
\left( 3 \hat \rho_{ab}^\alpha \hat \rho_{ab}^\beta - \delta_{\alpha \beta}
\right) , \nonumber \ \ \ \ \ \ \ \ \ \
\end{eqnarray}
where $h_{ab}^{\alpha \beta}=-\Delta_a^\beta \Delta_b^\alpha /J_\gamma$
and $h_{ab}^{\alpha \alpha}=\Delta_a^\beta \Delta_b^\beta /J_\gamma+
\Delta_a^\gamma \Delta_b^\gamma /J_\beta$ (all the three variables $\alpha$,
$\beta$ and $\gamma$ take different values here) denote the constant
quantities which are defined by the molecular geometry and $\rho_{ab}$ is
the distance between sites $a$ and $b$ within the same molecule.

Thus, the static Kirkwood factor of second order is uniquely determined by
the temperature and geometry of molecules. It is the same for infinite and
finite systems, i.e, $C_{\rm L,T}(k)=c \scs{\rm L,T}(k)$ (infiniteness of
the system has not been used at the derivation of equations (A4) and (A5)).
As can been shown earlier, this statement is completely in line with
transformations (9).
}

\newpage
\begin{center}
{\large Figure captions}
\end{center}

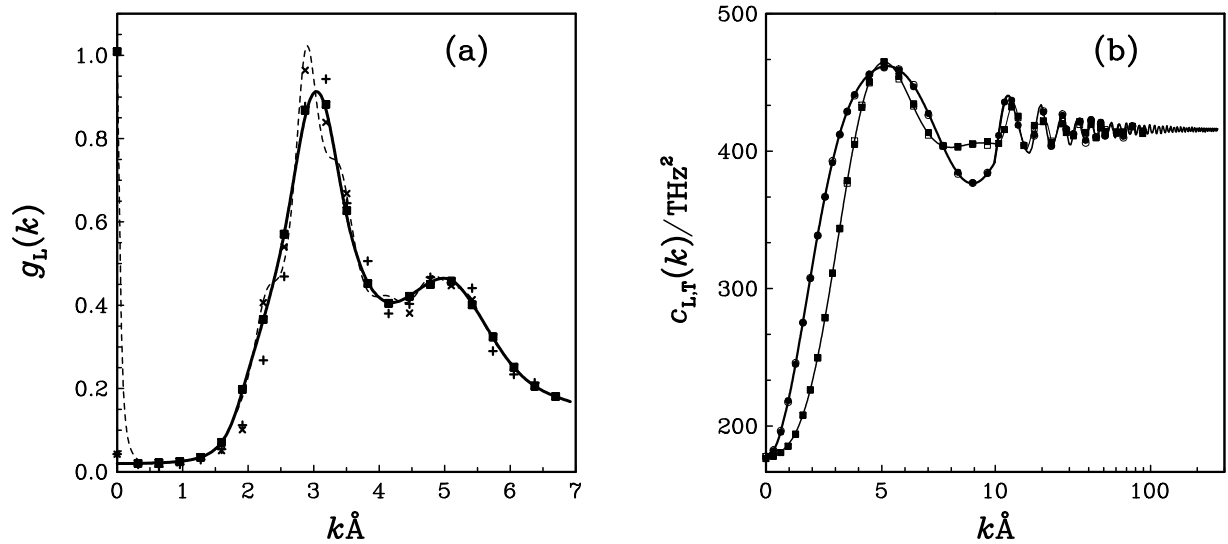
{\bf Fig.~1.}~The wavevector-dependent longitudinal Kirkwood factor $g
\scs{\rm L}(k)$ {\bf (a)} and the longitudinal and transverse components
$c \scs{\rm L,T}(k)$ {\bf (b)} of the current autocorrelation function for
the TIP4P water. The dashed and solid curves in {\bf (a)} correspond to the
finite and infinite systems in the ISRF geometry. The results of the Ewald
geometry for the longitudinal and transverse components are presented by the
full squares and circles, respectively, whereas the corresponding data in
{\bf (b)} obtained within the ISRF geometry are displayed as open squares
and circles. The analytical evaluation of $c \scs{\rm L,T}(k)$ (equation
(A5)) is plotted in {\bf (b)} by the solid curves. The calculations of
$g \scs{\rm L}(k)$, performed through the dynamical current correlations
(see equation (8)), are shown by the direct and rotated crosses for the
Ewald and ISRF geometries, respectively.

\vspace{20pt}

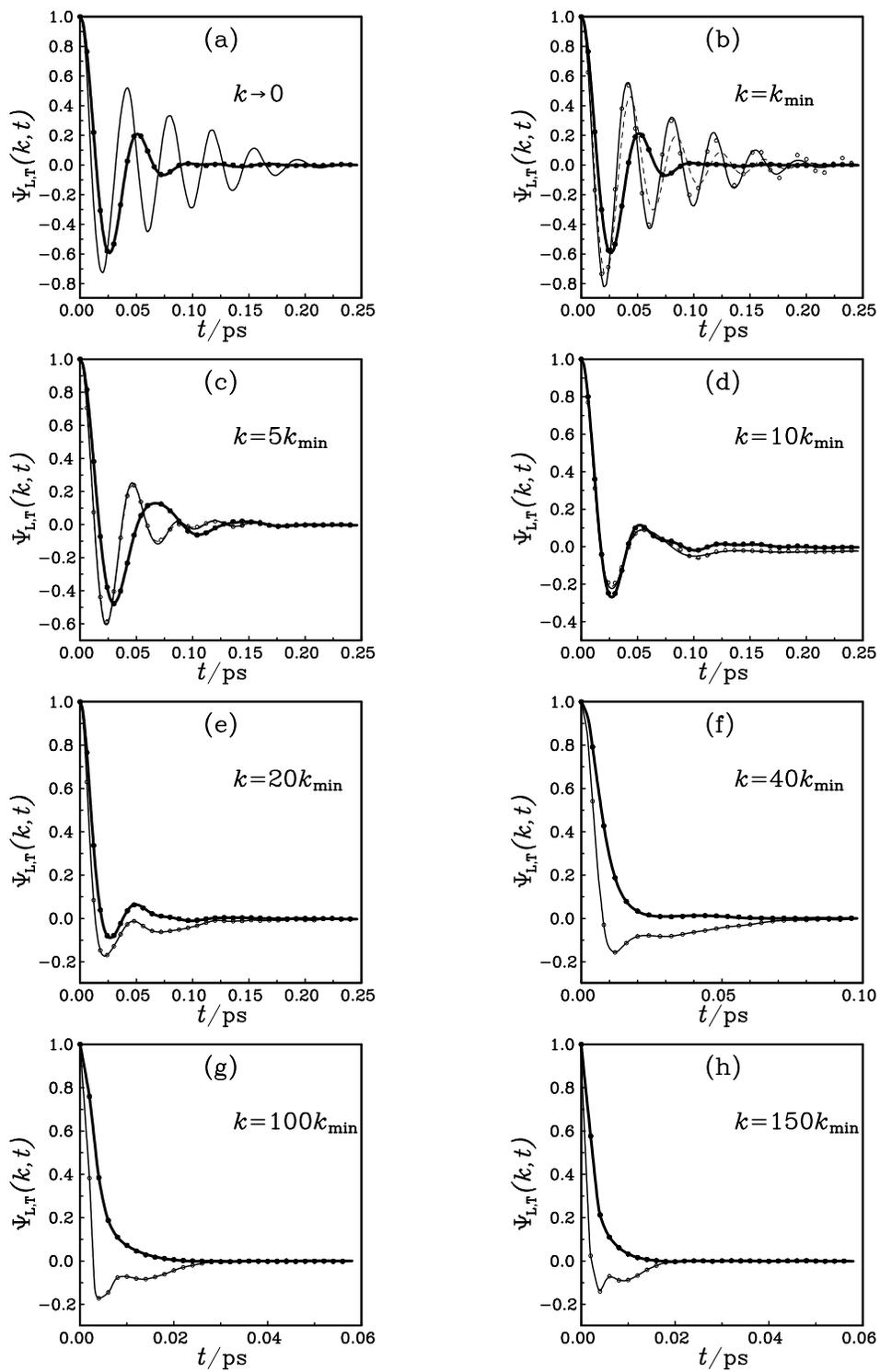
{\bf Fig.~2.} The normalized dynamical total current correlation functions
of the TIP4P water. The result of the Ewald geometry for the longitudinal and
transverse components is presented as open and full circles, respectively.
The data, obtained within the ISRF geometry for the infinite system, are
plotted by the corresponding solid curves. The longitudinal component
of the finite system within the ISRF geometry is shown as the dashed
curve.

\vspace{20pt}

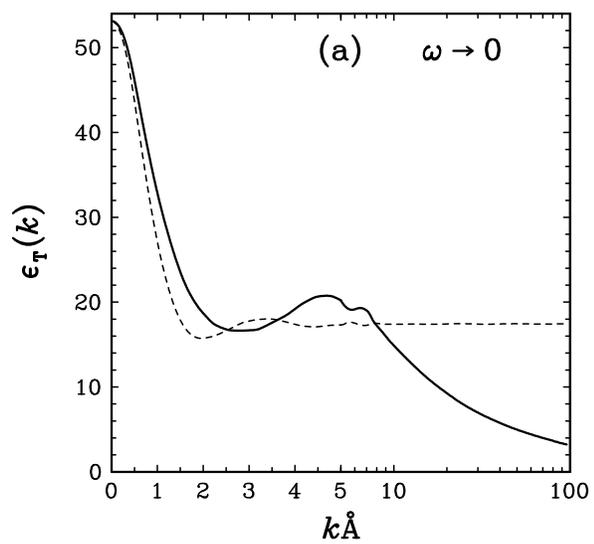
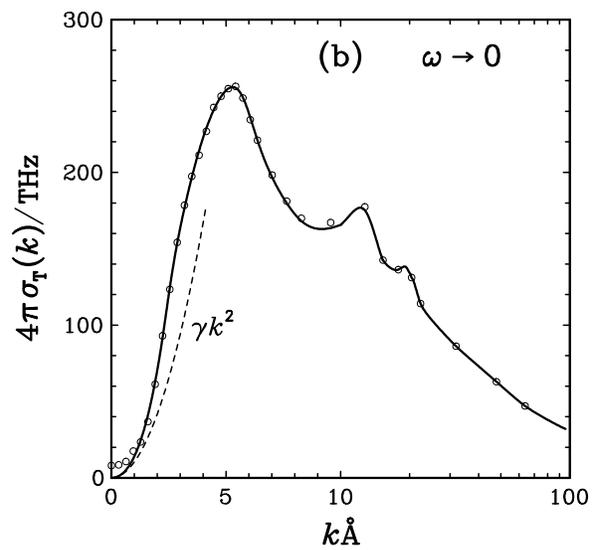
{\bf Fig.~3.} The wavevector-dependent transverse dielectric function
$\epsilon \scs{\rm T}(k)$ {\bf (a)} and generalized conductivity $\sigma
\scs{\rm T}(k)$ {\bf (b)} in the low frequency limit (the solid curves).
The dielectric function corresponding to the PD approximation and the
conductivity, calculated in terms of total current (instead of
magnetization) correlations are shown in subsets {\bf (a)} and
{\bf (b)} as the dashed curve and open circles, respectively.

\vspace{20pt}

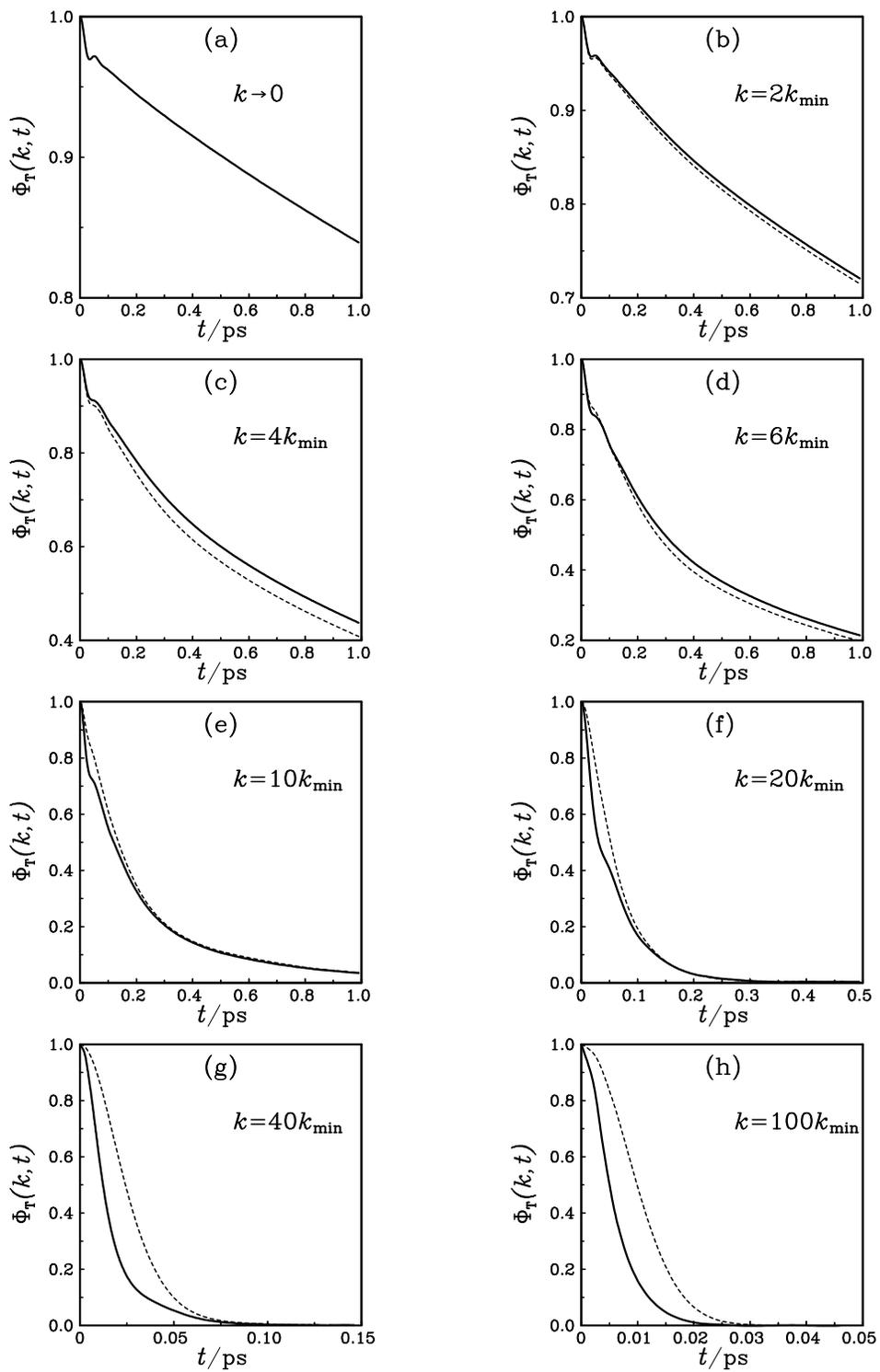
{\bf Fig.~4.} The normalized, time autocorrelation functions of the
transverse polarization fluctuations in the TIP4P water. The results
of the PD approximation are shown as dashed curves. Note that in the
infinite-wavelength limit $k\to0$ (see subset {\bf (a)}), the PD
functions are identical to genuine ones.

\vspace{20pt}

{\bf Fig.~5.} The normalized, time autocorrelation functions of the
transverse magnetization fluctuations in the TIP4P water. Notations
as for fig.~4.

\vspace{20pt}

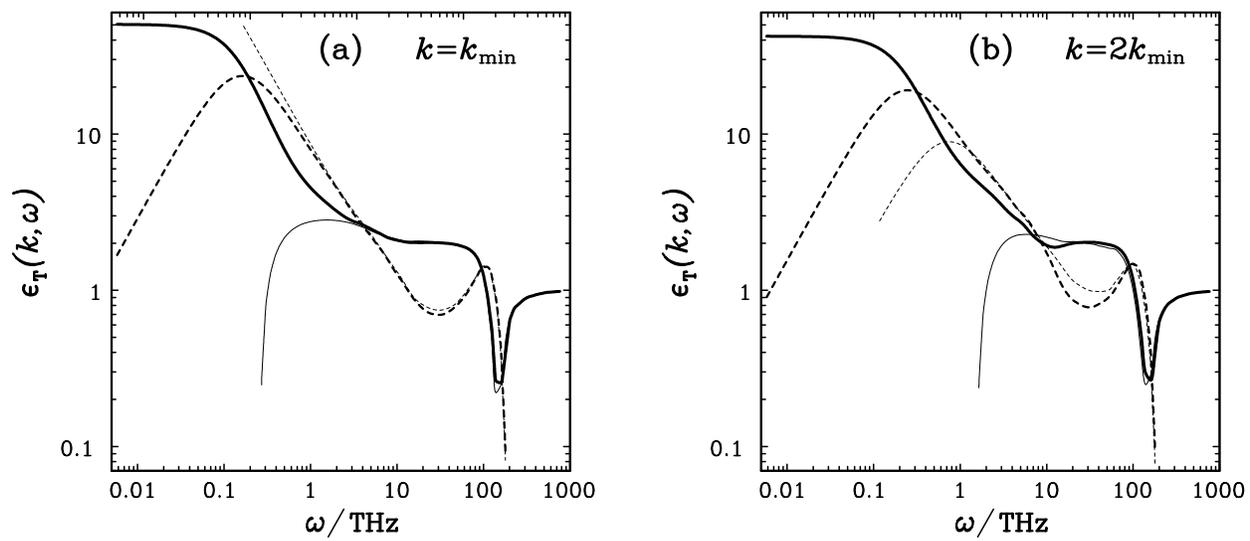
{\bf Fig.~6.} The transverse component of the wavevector- and
frequency-dependent dielectric function $\epsilon \scs{\rm T}(k,\omega)$
of the TIP4P water at small wavenumbers. The real and imaginary parts are
plotted by the bold solid and dashed curves, respectively. For comparison
the longitudinal component is shown by the corresponding thin solid and
dashed curves.

\vspace{20pt}

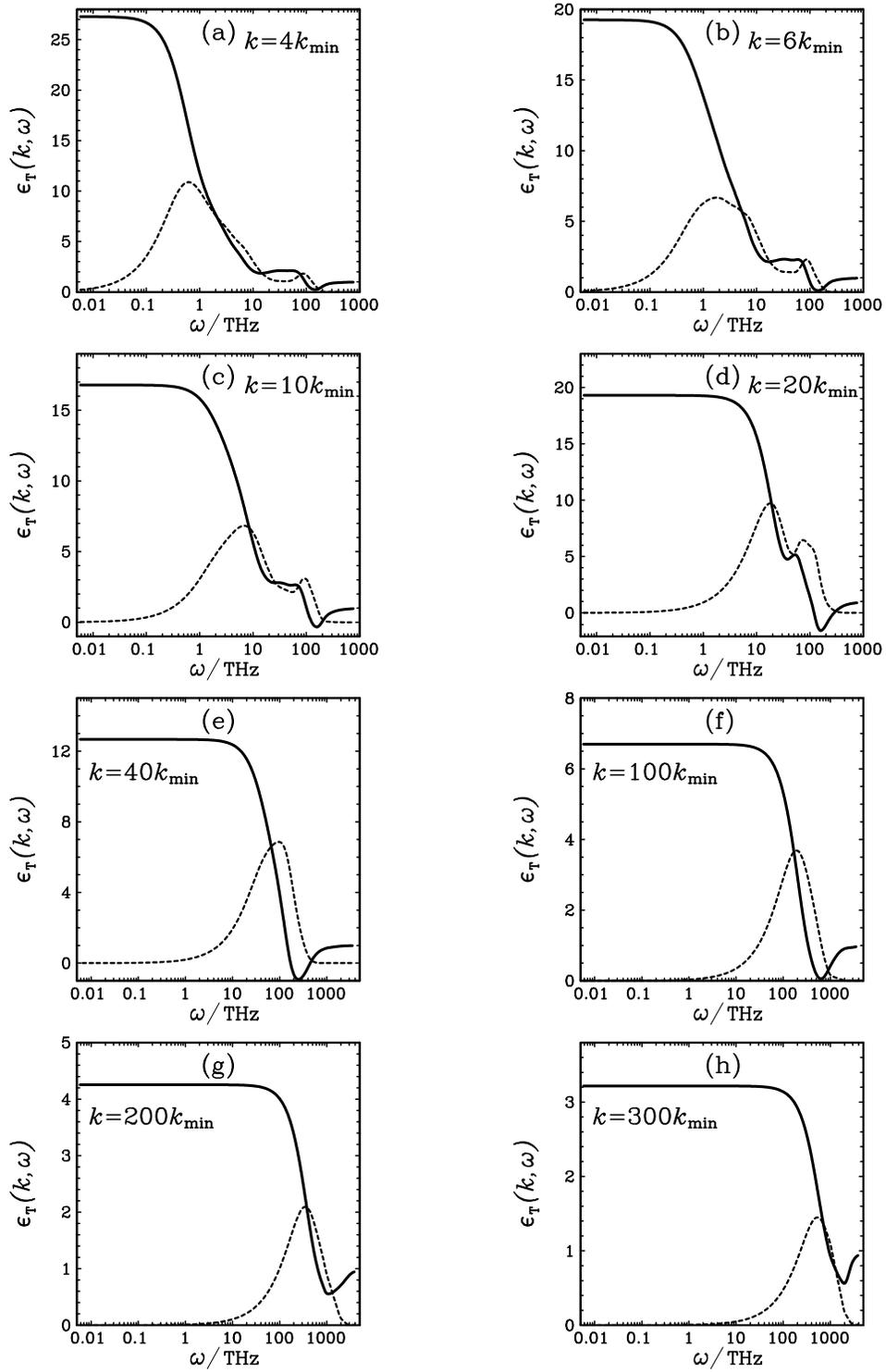
{\bf Fig.~7.} The transverse component of the wavevector- and
frequency-dependent dielectric function of the TIP4P water at great
wavenumbers. The real and imaginary parts are plotted by the solid and
dashed curves, respectively.

\vspace{20pt}

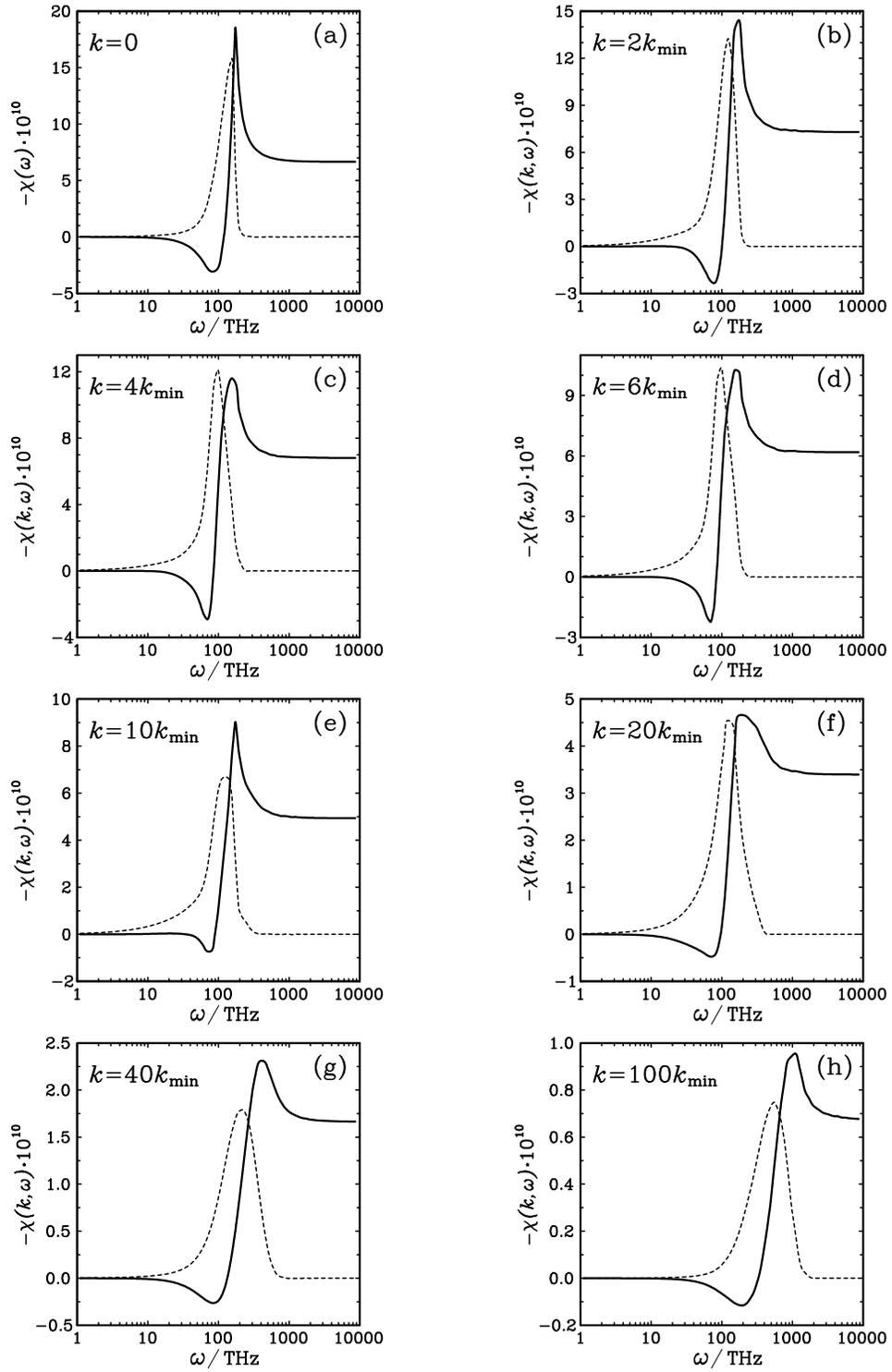
{\bf Fig.~8.} The wavevector- and frequency-dependent magnetic
susceptibility of the TIP4P water. The real and imaginary parts
are plotted by the solid and dashed curves, respectively. Note
that the results are shown in a negative representation.

\vspace{20pt}

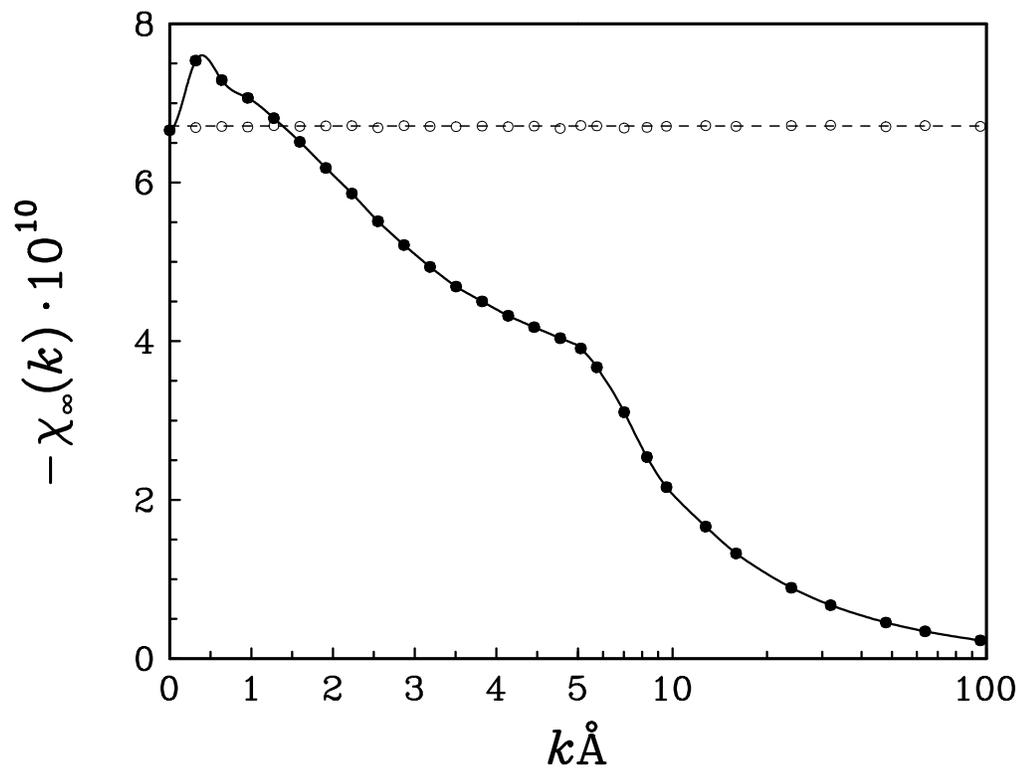
{\bf Fig.~9.} The wavevector-dependent magnetic susceptibility in the
infinite-frequency regime (full circles connected by the solid curve).
Note that the result in the PD approximation is independent on
wavenumbers (horizontal dashed curve). The PD function calculated
in the MD simulations is shown as open circles.

\vspace{20pt}

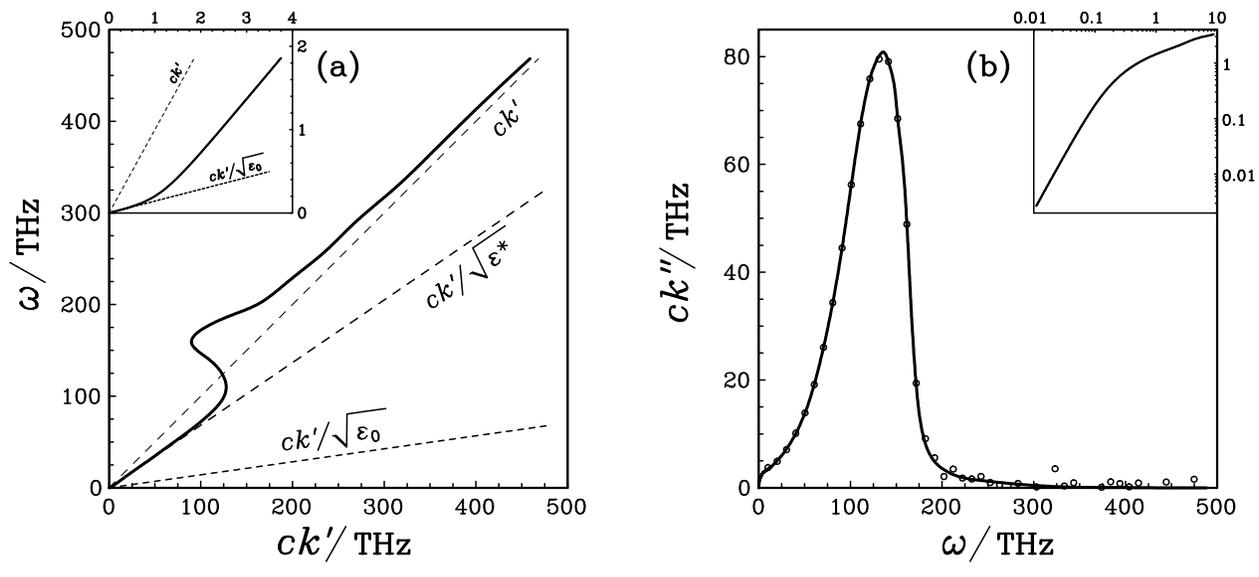
{\bf Fig.~10.} The dispersion {\bf (a)} and spatial damping {\bf (b)} of
transverse electromagnetic waves in the TIP4P water.

\vspace{20pt}

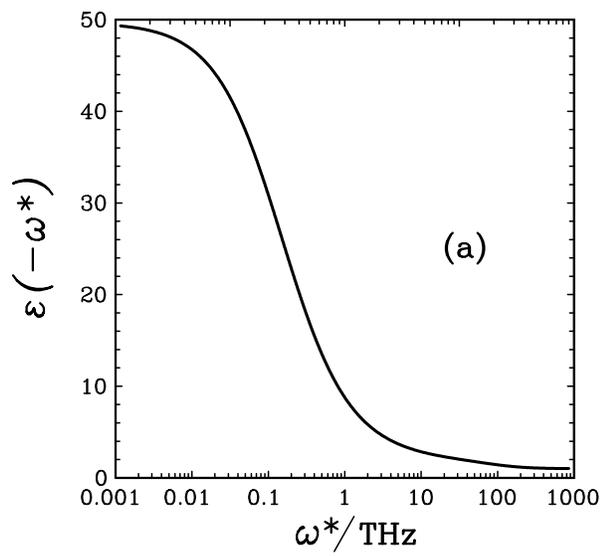
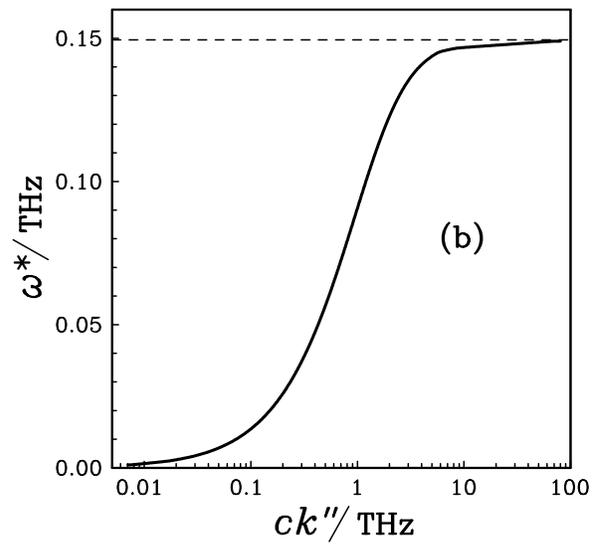
{\bf Fig.~11.} {\bf (a)} The dielectric constant of the TIP4P water in the
infinite-wavelength limit as a function of purely imaginary frequencies.
{\bf (b)} The time decay of electromagnetic fields in the system after
the passage of transverse electromagnetic waves.

\end{document}